\pdfoutput=1
\documentclass[11pt,preprint]{aastex}
\usepackage{graphicx}

\let\oldfootsep=\footnotesep
\newcommand\ltsima{$\; \buildrel <\over\sim \;$}
\newcommand\simlt{\lower.5ex\hbox{\ltsima}}
\newcommand\gtsima{$\; \buildrel >\over\sim \;$}
\newcommand\simgt{\lower.5ex\hbox{\gtsima}}

\newcommand\msun {M_\odot}

\newcommand\mearth {{M_\oplus}}

\newcommand\vp {\tilde{v}}
\newcommand\vpbold{\tilde{\bf v}}

\newcommand\pac{Paczy{\'n}ski }
\newcommand\ie{{i.e.}}

\newcommand{\mathbold}[1]{\mbox{\boldmath $\bf#1$}}
\newcommand\piEbold{{\mathbold \pi_E}}
\newcommand\rep {\tilde{r}_E}
\newcommand\repbold {{\mathbold \rep}}
 
\shorttitle{}
\shortauthors{Bennett et al}

\begin{document}


\title{A Low-Mass Planet with a Possible Sub-Stellar-Mass
         Host in Microlensing Event MOA-2007-BLG-192}


\author{D.P.~Bennett\altaffilmark{1,3},
               I.A.~Bond\altaffilmark{1,4},
               A.~Udalski\altaffilmark{2,5},
               T.~Sumi\altaffilmark{1,6},
               F.~Abe\altaffilmark{1,6}, 
               A.~Fukui \altaffilmark{1,6}, 
               K.~Furusawa\altaffilmark{1,6}, 
               J.B.~Hearnshaw\altaffilmark{1,7},
               S.~Holderness\altaffilmark{1,8},
               Y.~Itow\altaffilmark{1,6}, 
               K. Kamiya\altaffilmark{1,6},
               A.V.~Korpela\altaffilmark{1,9},
               P.M.~Kilmartin\altaffilmark{1,10},
               W.~Lin\altaffilmark{1,4},
               C.H.~Ling\altaffilmark{1,4},
               K.~Masuda\altaffilmark{1,6}, 
               Y.~Matsubara\altaffilmark{1,6}, 
               N.~Miyake\altaffilmark{1,6},
               Y.~Muraki\altaffilmark{1,11}, 
               M.~Nagaya\altaffilmark{1,6},
               T.~Okumura\altaffilmark{1,6}, 
               K.~Ohnishi\altaffilmark{1,12}, 
               Y.C.~Perrott\altaffilmark{1,13}, 
               N.J.~Rattenbury\altaffilmark{1,14}, 
               T.~Sako\altaffilmark{1,6}, 
               To.~Saito\altaffilmark{1,15},
               S.~Sato\altaffilmark{1,16}, 
               L.~Skuljan\altaffilmark{1,4},
               D.J.~Sullivan\altaffilmark{1,9}, 
               W.L.~Sweatman\altaffilmark{1,4},
               P.J.~Tristram\altaffilmark{1,10},
               P.C.M.~Yock\altaffilmark{1,13},
               M.~Kubiak\altaffilmark{2,5}, 
               M.K.~Szyma{\'n}ski\altaffilmark{2,5}, 
               G.~Pietrzy{\'n}ski\altaffilmark{2,5,17}, 
               I.~Soszy{\'n}ski\altaffilmark{2,5},
               O.~Szewczyk\altaffilmark{2,5,17}, 
               {\L}.~Wyrzykowski\altaffilmark{2,5,18}, 
               K.~Ulaczyk\altaffilmark{2,5}, 
               V.~Batista\altaffilmark{19},
               J.P.~Beaulieu\altaffilmark{19},
               S.~Brillant\altaffilmark{20},
               A.~Cassan\altaffilmark{21},
               P.~Fouqu\'e\altaffilmark{22},
               P. Kervella\altaffilmark{23},
               D.~Kubas\altaffilmark{20}, and
               J.B.~Marquette\altaffilmark{19}
              } 


\begin{abstract}
We report the detection of an extrasolar planet of mass ratio 
$q \sim 2 \times 10^{-4}$ in microlensing event MOA-2007-BLG-192.
The best fit microlensing model shows both the microlensing parallax 
and finite source effects, and these can be combined to obtain the lens
masses of $M = 0.060 {+0.028\atop -0.021}\msun$ for the primary and 
$m = 3.3{+4.9\atop -1.6} \mearth$ for the planet. However, the observational 
coverage of the planetary deviation is sparse and incomplete, and the radius
of the source was estimated without the benefit of a source star color 
measurement. As a result,
the 2-$\sigma$ limits on the mass ratio and finite source
measurements are weak. Nevertheless, 
the microlensing parallax signal clearly favors a sub-stellar mass planetary
host, and the measurement of finite source effects in the light curve 
supports this conclusion. Adaptive
optics images taken with the Very Large Telescope (VLT) NACO instrument 
are consistent with a lens star that is either a brown dwarf or a star at the 
bottom of the main sequence. Follow-up VLT and/or Hubble Space 
Telescope (HST) observations will either confirm that the primary is
a brown dwarf or detect the low-mass lens star and enable a precise 
determination of its mass. In either case, the lens star, MOA-2007-BLG-192L,
is the lowest mass primary known to have a companion with a planetary
mass ratio, and the planet, MOA-2007-BLG-192Lb, is probably the lowest mass
exoplanet found to date, aside from the lowest mass pulsar planet.
\end{abstract}



\clearpage


\altaffiltext{1}{MOA Collaboration}

\altaffiltext{2}{OGLE Collaboration}

\altaffiltext{3}{Department of Physics,
    University of Notre Dame, IN 46556, USA\\
    Email: {\tt bennett@nd.edu}}

\altaffiltext{4}
{Institute of Information and Mathematical Sciences, Massey University,
Auckland, New Zealand,\\
Email: {\tt i.a.bond@massey.ac.nz}}

\altaffiltext{5}{Warsaw University Observatory,
    Al. Ujazdowskie 4
    00-478 Warszawa, Poland\\
    Email: {\tt udalski@astrouw.edu.pl}}

\altaffiltext{6}{Solar-Terrestrial Environment Laboratory, 
    Nagoya University, Nagoya 464-8601, Japan\\
    Email: {\tt sumi@stelab.nagoya-u.ac.jp}}
    
\altaffiltext{7}{Department of Physics and Astronomy, 
    University of Canterbury, Private Bag 4800, 
   Christchurch, New Zealand}

\altaffiltext{8}{Computer Science Department, University of Auckland, 
     Auckland, New Zealand}
     
\altaffiltext{9}{School of Chemical and Physical Sciences, 
     Victoria University, Wellington, New Zealand}
 
\altaffiltext{10}{Mt. John Observatory, P.O. Box 56,
      Lake Tekapo 8770, New Zealand}

\altaffiltext{11}{Konan University, Kobe, Japan}

\altaffiltext{12}{Nagano National College of Technology, 
      Nagano 381-8550, Japan}

\altaffiltext{13}{Department of Physics, University of Auckland, 
     Auckland, New Zealand}
     
\altaffiltext{14}{Jodrell Bank Observatory, The University of 
      Manchester, Maccles eld, Cheshire SK11 9DL, UK}
      
\altaffiltext{15}{Tokyo Metropolitan College of Aeronautics, 
      Tokyo 116-8523, Japan}
      
\altaffiltext{16}{Department of Physics and Astrophysics, 
     Faculty of Science, Nagoya University, Nagoya 464-8602, Japan}

\altaffiltext{17}{Departamento de Fisica, Astronomy Group, Universidad de
Concepci\'on, Casilla 160-C, Concepci\'on, Chile}

\altaffiltext{18}{Institute of Astronomy, University of Cambridge, 
      Madingley Road, Cambridge CB3 0HA, UK}

\altaffiltext{19}{Institut d'Astrophysique de Paris, UMR7095 CNRS,
    Universit\'{e} Pierre \& Marie Curie, 98~bis Boulevard Arago, 75014 Paris, France}
      
\altaffiltext{20}{European Southern Observatory, Casilla 19001, 
     Vitacura 19, Santiago, Chile}
      
\altaffiltext{21}{Astronomisches Rechen-Institut, Zentrum f\"ur~Astronomie, Heidelberg
      University, M\"{o}nchhofstr.~12--14,69120 Heidelberg, Germany}

             
\altaffiltext{22}{Observatoire Midi-Pyr\'en\'ees, UMR 5572, 14, 
         Avenue Edouard Belin, 31400 Toulouse, France}

\altaffiltext{23}{LESIA, Observatoire de Paris, CNRS UMR 8109, UPMC,
         Universit\'e Paris Diderot, 5 Place Jules Janssen,
         F-92195 Meudon, France}
      
\section{Introduction}
When the first extrasolar planets were discovered orbiting main sequence
stars more than a decade ago \citep{51peg,70vir,47uma}, the radial velocity surveys responsible
for the discoveries focused their observations on stars of spectral type
F, G, and K, because such stars offered the greatest planet detection
sensitivity. However, observations of star forming regions indicate
that stars of virtually all types have evidence of proto-planetary disks,
suggesting that the first stages of the planet formation are
practically independent of star type. The radial velocity surveys have
since expanded their target lists to include stars ranging from a
spectral type of mid-M to stars that are thought to have spectral
type A when they were on the main sequence. Gas giant planets
have been found in orbit around stars of all these types
\citep{butler-catalog,sato_plan_giant,johnson_Mdwarf_plan,psu_Gstar_pl,lovis_pl_evolved},
but they appear to be increasingly rare in orbit around low-mass
stars.

In the past few years, microlensing surveys 
\citep{exoplanet_book} have extended the
range of sensitivity to cool,  ``super-Earth" planets 
\citep{ogle390,ogle169} with
masses of $\sim 10\mearth$ in orbits beyond the ``snow-line"
\citep{ida_lin,laughlin,kennedy-searth}, where the core accretion
theory predicts that the most massive planets should form. 
These discoveries indicate that such low-mass
planets are significantly more common than gas giants in orbit
around  the stars of $\simlt 1\msun$ that are probed by the 
microlensing method. Because microlensing does not rely
upon light from the planetary host star in order to detect
the planet, its sensitivity extends to host star masses 
well below the $\sim 0.25\msun$ lower limit for current
radial velocity surveys. 

In this paper, we present the analysis of microlensing
event MOA-2007-BLG-192, and show that 
the lens system is likely to be a low-mass planet
orbiting a primary that is either a brown dwarf or
a very low-mass main sequence star. The data are
discussed in \S~\ref{sec-data}, and the planetary
nature of the light curve is discussed in \S~\ref{sec-lc}.
The uncertainties in the microlensing model parameters are 
discussed in \S~\ref{sec-param_un}, and in
\S~\ref{sec-lens_prop}, we show that the
microlensing parallax and finite source features of the light
curve favor a sub-stellar mass primary and a very low-mass
planet. Adaptive optics images from the VLT NACO instrument
are only consistent with a primary lens mass that is a brown dwarf or
a star at the bottom of the main sequence. In \S~\ref{sec-conclude},
we show how future observations with VLT/NACO can
confirm this interpretation and determine the
parameters of the planetary system more precisely.

\section{Data}
\label{sec-data}
The discovery of 
microlensing event MOA-2007-BLG-192 was triggered by data from 
the night of peak magnification, 2007 May 24 UT, when
planetary deviation occurred. The faintness of the source
star and poor weather at the MOA telescope during the
night prior to peak magnification prevented
earlier detection of this event by MOA. 

The 2.2 square degree field of the MOA-cam3,
mounted on the the 1.8m MOA-II telescope \citep{sako_moacam3,moa2_tel}
allows 50 square degrees of the Galactic bulge to be 
observed every hour, and it was this frequent sampling of
this event that resulted in the detection of the planetary signal,
despite the lack of an early alert. Continuing improvements
in the MOA photometry and alert system \citep{moa-alert} should allow a 
similar event to be alerted earlier in future seasons.

The photometry of the MOA data was performed with a
custom version of the MOA pipeline that is optimized for
precise photometry of selected events. 
When an event is detected, a collection of small
$256 \times 256$ pixel ``cameo" images is generated from 
all the images taken that season. These
small images provide more precise
coordinate transformations and cleaner subtracted images,
resulting in more precise photometry. 
We selected the best MOA photometry from multiple runs with
different photometry code parameters by comparing light
curve fits to the data outside of the planetary deviation.
The photometry of the light curve peak, which is 
critical for the planetary interpretation was confirmed by 
multiple photometry codes including the
OGLE pipeline \citep{ogle-pipeline}.
The MOA data set consists of 718 observations from
the 2007 season.

This event was not among the $\sim 600$ events per year found
by the OGLE Early Warning System
\citep{ogle-ews} because the source star was too faint to appear
in the OGLE star catalog used for online photometry. 
However, a star catalog based upon
a recent OGLE image with 0.75" seeing does include the
MOA-2007-BLG-192 source star.
The OGLE photometry was obtained
with the standard OGLE photometry pipeline with this good seeing
reference image, and the resulting
data set consists of 442 observations dating back to 2001.

In the crowded stellar fields where microlensing events are 
observed, the true photometric errors often depend on the
proximity of nearby stars. As a result, it is 
customary to rescale the error bars to give $\chi^2/{\rm d.o.f.} \simlt 1$
for each telescope/passband. For this event, these scaling factors
were determined with a single lens fit excluding the data taken in
a 24-hour period centered on the planetary light curve deviation.
We find that no rescaling is needed for the MOA data, while the
OGLE error bars are slightly overestimated. Both data sets have
a 0.1\% systematic uncertainty added in quadrature, and the reported
OGLE error bars are reduced by a factor of 0.92.

\section{Planetary Nature of the Light Curve}
\label{sec-lc}

This event is an example of a high magnification event, which allows
the detection of planets via perturbations of the central or ``stellar" caustic
\citep{griest_saf,mps-98blg35,ratt,ogle71,ogle169,gaudi-ogle109}. 
But close or wide stellar binary lens
systems can also give light curve perturbations at high magnification
\citep{planet_m99b47,moa2002blg33}, so it is important to distinguish 
between these
two possibilities. Also, the incomplete coverage of the light curve may
allow multiple binary lens models, so it is important to do a careful 
search of parameter space to ensure that all viable models are
found.

\begin{figure}
 \epsscale{0.90}
\plotone{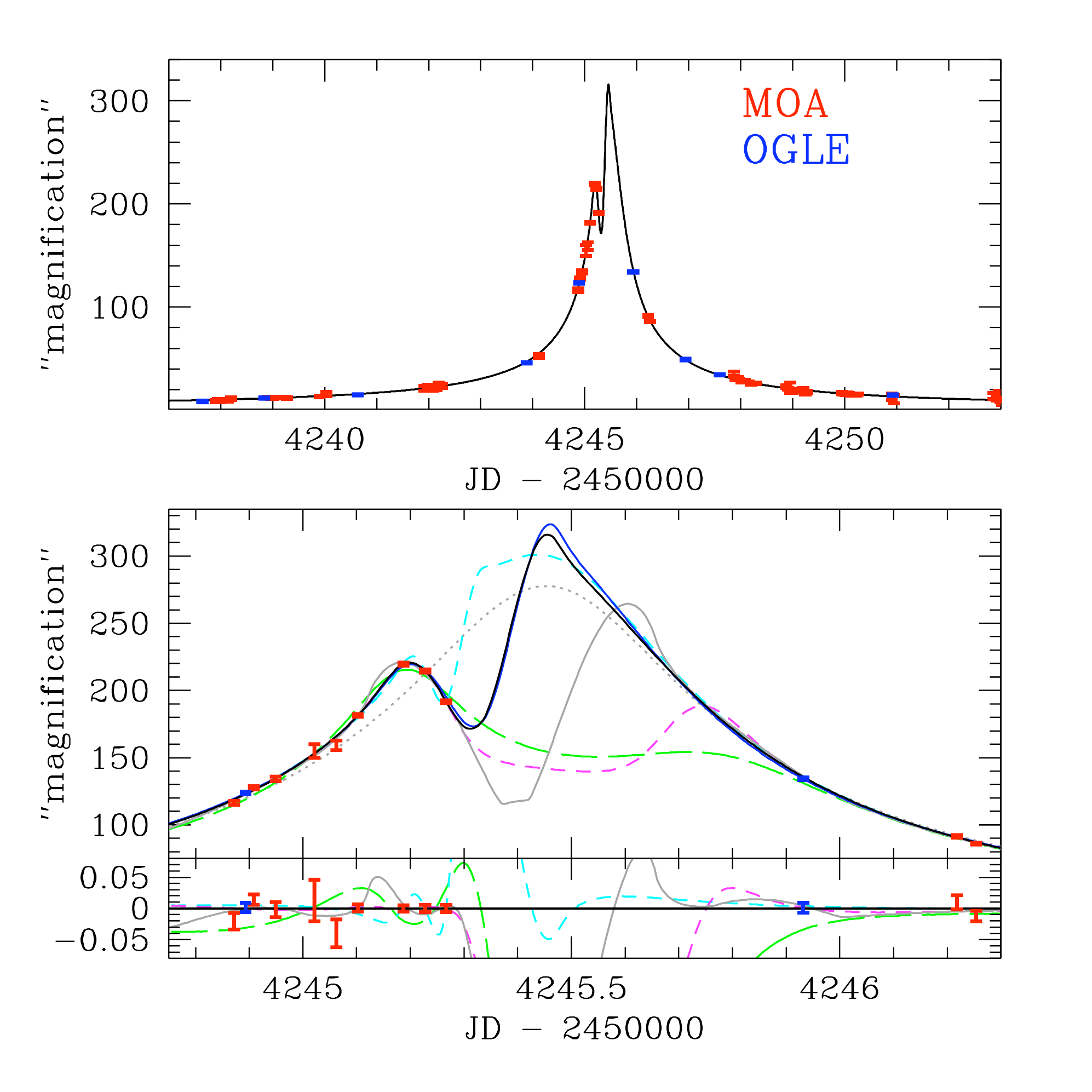}
\caption{
The microlensed portion of the light curve of MOA-2007-BLG-192 as
seen by the MOA telescope
is plotted in flux units normalized to the $I_s = 21.48$ source of
model B from Table~\ref{tab-fitpar} The light curve peak is shown on the
large lower panel and the 
bottom sub-panel shows the fractional deviation of the data
from the best fit model, which is indicated by the solid black curve
in both panels. A number of alternative light curve models are shown
in addition to the best fit model. 
The solid grey curve is the best caustic crossing model, and
the short dashed cyan and magenta curves are the models corresponding
to the 2-$\sigma$ lower and upper limits on the mass ratio, $q$.
The dotted grey curve 
is the single lens model with the same parameters as the best 
fit model, and the long-dashed green curve is the best fit stellar binary model.
MOA and OGLE data are plotted in red and blue, respectively. The
solid blue curve is the best fit model for the OGLE data, which differs
from the black (MOA) curve due to terrestrial parallax. The black/blue and
grey curves represent models B and I of Table~\ref{tab-fitpar}, respectively.
\label{fig-lc}}
\end{figure}

We have carefully searched parameter space to find the
best fit planetary and stellar binary lens models, and these best fit models
are displayed in Figure~\ref{fig-lc} with the solid black curve indicating the
best fit planetary model and the green curve in the right hand panel
indicating the best fit stellar binary model. 
Figure~\ref{fig-lc} shows the magnified portion of the MOA-2007-BLG-192
light curve, with a close-up of the light curve peak shown in the lower
panel. These plots are made in linear flux units normalized to the flux
of the best fit planetary lens model, which has a source magnitude of
$I_s = 21.48$. The black and blue curves indicate the best fit model
as seen from the MOA and OGLE telescopes, respectively. These
light curves are slightly different due to the different microlensing
magnification observed from the sites of the different telescopes.
This effect is known as terrestrial parallax and was first discussed by
\citet{holz_wald}. For all the other models, only the light curve as
seen by MOA is shown.

The grey curve is a caustic crossing planetary model, which has a similar
$\chi^2$ and planetary parameters to the best fit planetary model, but 
has a very different caustic structure. In fact, the fit $\chi^2$ is slightly
lower than the ``best" model, but this slight $\chi^2$ improvement is
more than offset by the lower {\it a priori} probability of the parameters.
The short-dashed cyan and 
magenta curves represent the 2-$\sigma$ lower and upper limits on
the planetary mass ratio, $q$, so they have a $\chi^2$ value that is larger
than the best model by $\Delta\chi^2 = 4$. The reasons why these models
provide the lower and upper mass ratio limits are clear from the light
curve plot. The lower mass ratio limit light curve (cyan) puts the last
MOA observation on the planetary deviation at the minimum between
the two cusp approach peaks. At smaller $q$, the separation between
these cusps would grow smaller, which would tend to increase the
brightness at the time of this observation ($t = 4245.27$). The upper
limit model (magenta) has a much larger cusp approach separation and 
has pushed the second cusp approach to $t \approx 4245.75$ where it
begins to affect the OGLE observation at $t = 4245.93$. Thus, the planetary
models are constrained to have the second cusp approach appear
in the time interval $4245.3 < t < 4245.75$ where it is not constrained
by the data. Similar arguments can be made with regard to the caustic crossing
model shown by the grey curve, but the limits on $q$ for the caustic
crossing models are tighter than the limits on the cusp approach models.

The long-dashed green curve is the best fit stellar binary model, which is
able to reproduce the observed light curve to within a few percent. But, it 
does not provide a good fit to the data. The best fit model has 
$\chi^2 = 1115.46$ for 1160 data points and 13 model parameters
to give $\chi^2/{\rm d.o.f.} = 0.9729$, while the best fit stellar binary
model gives $\chi^2 = 1237.40$, a difference of $\Delta\chi^2 = 121.94$.

\begin{figure}
\plotone{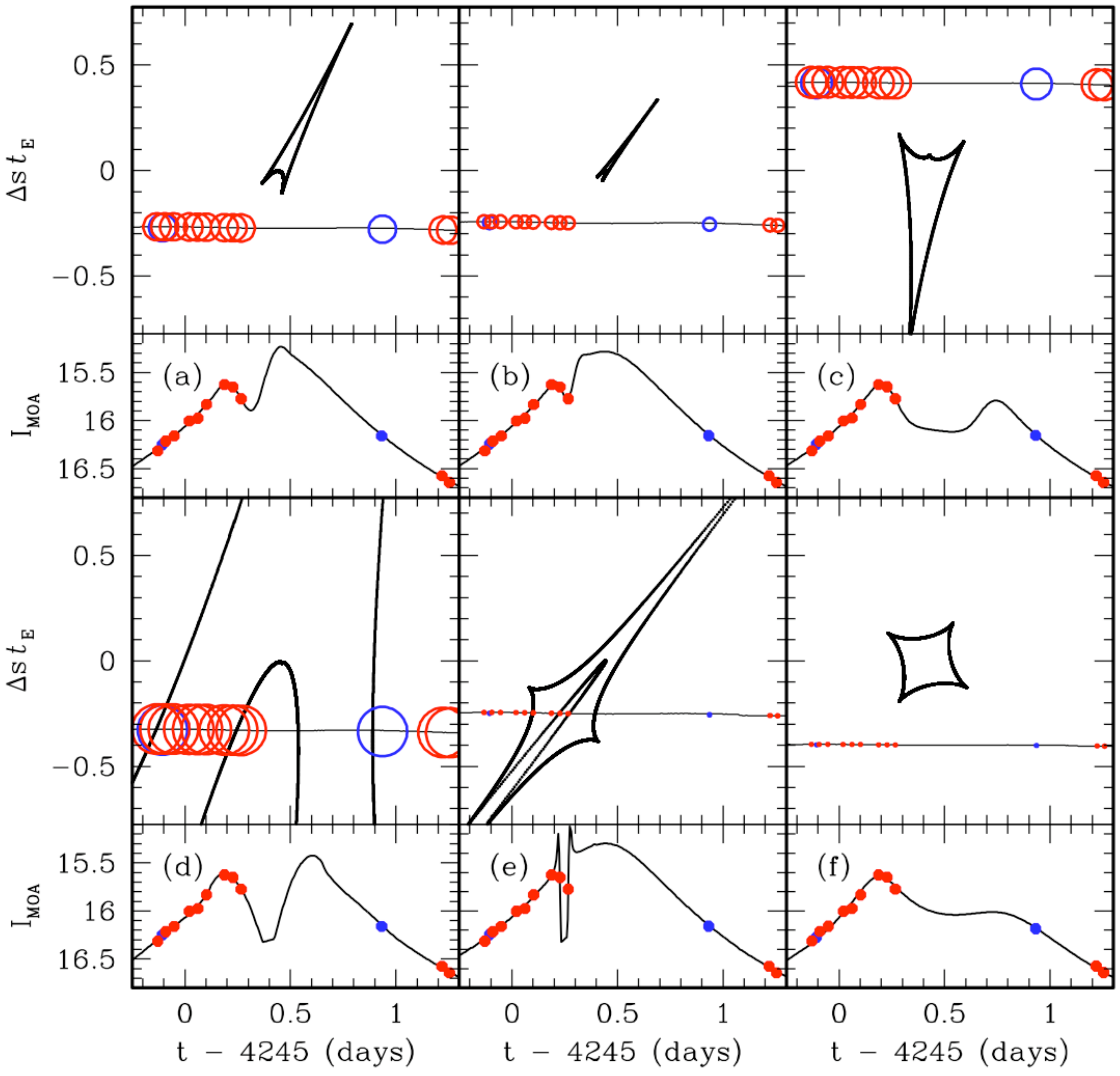}
\caption{
The caustic geometries for the 5 binary lens
fits shown in Fig.~\ref{fig-lc} are plotted. The lens source trajectories are given by
the horizontal lines, with the red (MOA) and blue (OGLE) circles indicating the
timing of the images and the relative size of the source. The model light curves
are shown in the lower sub-panels with the MOA and OGLE measurements
in red and blue.
(a) shows the best fit model, while (b) and (c) show the mass ratio 2-$\sigma$ 
lower and upper limit models. (d) shows the best caustic crossing model, while
(e) shows an alternative caustic crossing model that we reject on
{\it a priori} grounds. (f) shows the best fit stellar binary model, which does not
provide an acceptable fit to the data. Panels (a) and (d) correspond to models
B and I of Table~\ref{tab-fitpar}, respectively.
\label{fig-caustic}}
\end{figure}

The reason why the stellar binary models fail can be understood based
upon the basic properties of the central caustics of stellar binary
light curves, \ie\ those without an extreme mass ratio. The central
caustics of such binaries are quite strong, so they provide very
strong light curve deviations that extend far from the light curve
peak unless they are kept very small by
making the binary separation, $d$, very much smaller or very much
larger than the Einstein ring radius. This forces the caustics to
the diamond shaped form shown in Figure~\ref{fig-caustic}(f).

\begin{figure}
\plotone{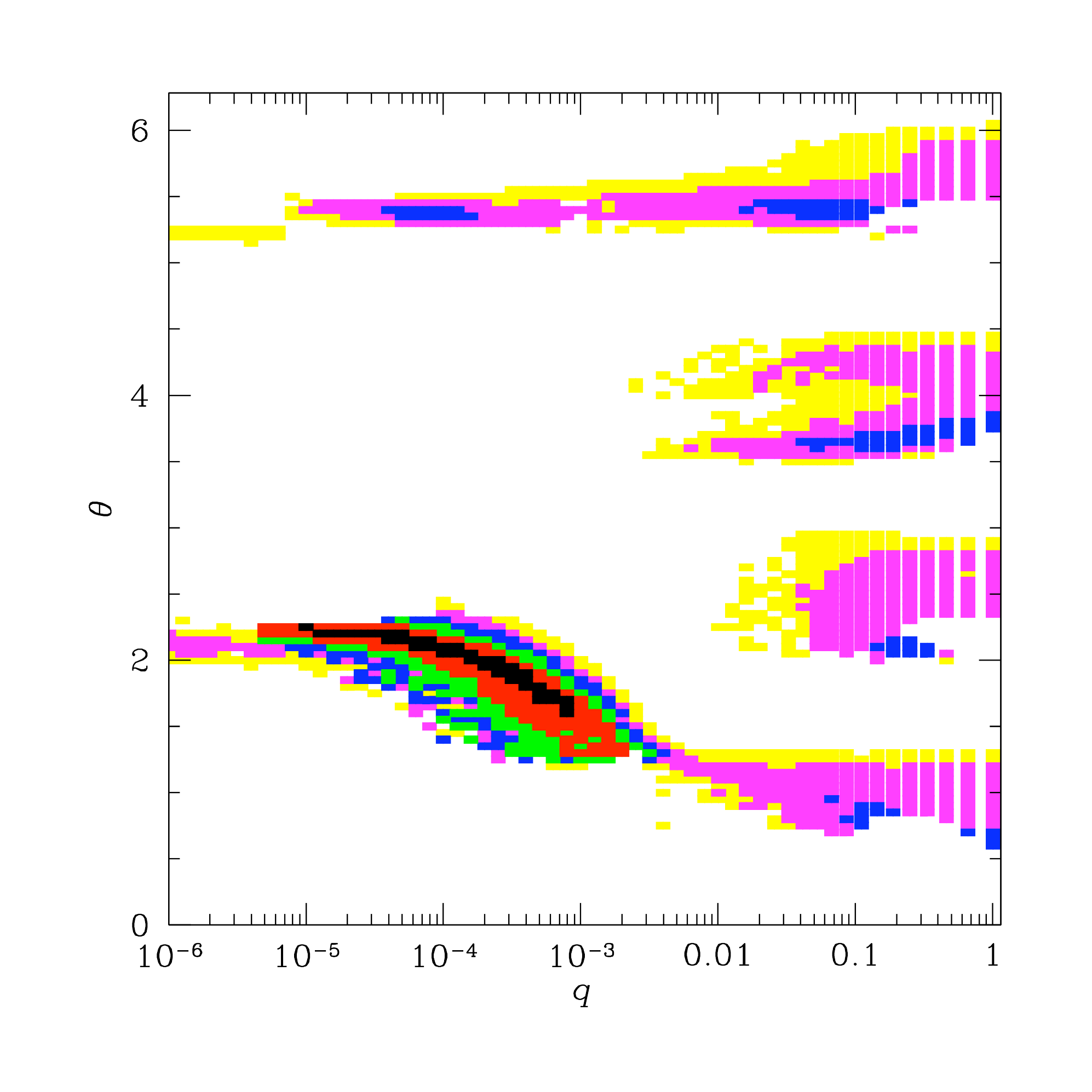}
\caption{The best binary lens fit $\chi^2$ values for fixed $q$ and
$\theta$ are indicated by the colored regions.
The black, red, green, blue, magenta, and yellow regions indicate the
areas of parameter space that are excluded by $\Delta\chi^2 = 10$, 40, 90,
160, 250, and 360, respectively. In the white regions, the best fit $\chi^2 > 360$.
For most regions of parameter space, we have used the point source
approximation, since a finite source cannot significantly improve the fit
$\chi^2$ over a point source model. The only exceptions are the regions
of planetary solutions with $q \simlt 10^{-3}$ and $\theta \sim 5.4$ or
$\theta \sim 1.9$. This figure includes a finite source for the 
$\theta \sim 5.4$ planetary models, but not for the $\theta \sim 1.9$, as the
later are studied in much greater detail later in this paper.
\label{fig-q_theta_grid}}
\end{figure}

Figure~\ref{fig-q_theta_grid} shows the parameter space locations  of 
the models that can provide an approximate fit to the light curve data
with a $\Delta\chi^2$ value within 360 of the best fit as a function of the
mass ratio, $q$, and the 
angle between the lens axis and the direction of the lens-source relative
motion, $\theta$. For stellar binary solutions, with $q \simgt 0.1$, this 
figure indicates an approximate 4-fold degeneracy, with the best
solutions at each $q$ value separated by $\Delta\theta \sim 90^\circ$.

The situation is different for the planetary models with $q \simlt 10^{-3}$.
Figure~\ref{fig-q_theta_grid} shows that only two of these solutions
continue to give moderately low $\chi^2$ values in the 
planetary regime. So, the approximate $90^\circ$ symmetry at large
$q$ has morphed into an approximate $180^\circ$ symmetry at small
$q$. This is easily understood as following from the basic properties
of the central caustics of binary lens light curves \citep{dominik99}.
The central caustic for the best fit stellar binary model is shown in
Figure~\ref{fig-caustic}(f). This model has a mass ratio of $q = 0.59$, but
the central caustic has a nearly perfect $90^\circ$ rotation symmetry.
The regions of higher magnification extend outward from the cusps
of these caustic curves, and so one of the cusps points at the location
of the source at $t = 4245.2$ in order to account of the observed
bump in the light curve. But, the $90^\circ$ symmetry enforces a
minimum $\sim 0.6\,$day delay between the observed cusp approach
and the next one. This ensures that the 2nd cusp approach will have
some effect on the OGLE data point at $t = 4245.93$.

Figures~\ref{fig-caustic}(a), (b), and (c) show the caustic curves
and source trajectories for the best fit and the  2-$\sigma$ lower
and upper limit (on $q$) models. The red and blue circles indicate
the location and relative size of the source star at the time of the 
MOA and OGLE images, respectively. The magnification deviations
due to these central caustics extend outward from the cusps, but the 
cusps in these planetary models always point in a direction quite close
to the lens axis. Thus, the two sets of planetary solutions shown in
Figure~\ref{fig-q_theta_grid} correspond to the cases where the
source crosses the single ``forward" cusp or approaches the multiple
cusps pointing ``backwards\rlap." (There are always three cusps pointing
``backwards" as shown in Figure~\ref{fig-caustic}(c), but when the planet
is close to the Einstein ring, as is the case for the small $q$ solutions,
the central caustic on the ``back" side becomes so weak as to be invisible
in these figures.)

The caustic geometries of two caustic crossing models are shown in
Figures~\ref{fig-caustic}(d) and (e). The model given by the grey curve in
Figure~\ref{fig-lc} corresponds to Figure~\ref{fig-caustic}(d), while 
the light curve corresponding to Figure~\ref{fig-caustic}(e) is not shown in
Figure~\ref{fig-lc}, although it is shown without the data points in the lower
sub-panel of Figure~\ref{fig-caustic}(e). This model has the distinction
of having 3 of the final 4 MOA observations on day 4245 just happen
to occur during 3 separate caustic crossings. Furthermore, the observations
during the two very strong caustic crossings just happen to have a very similar
magnitude to each other and the previous two observations. Since the 
magnification is changing very rapidly on these later two caustic crossings
there is a
very small {\it a priori} probability to obtain a light curve like the observed
one if a model like this is correct. The light curve spends much more time
at much higher and lower magnifications than at the observed magnifications.
To put this another way, if the model of Figure~\ref{fig-caustic}(e) is correct,
it is very unlikely to find a model with a smoother light curve like the
model of Figure~\ref{fig-caustic}(a) that also fits the data. But if the
model of Figure~\ref{fig-caustic}(a) is correct, then it is likely that we
can find a model like that of Figure~\ref{fig-caustic}(e) that also
fits the data because there are enough parameters in the model to
adjust the caustic crossing times to match the times of the relatively
sparse observations. It is also worth noting that the source sizes
for the (e) and (f) models are unphysically small, although there are
similar models with reasonable source sizes that have $\chi^2$ values
that are larger by $\Delta\chi^2 \approx  4$.

As a final check of the planetary nature of the light curve, we
have performed extensive comparisons of the best fit planetary
and stellar binary models with different photometry parameters and 
with some of the critical points removed from the data sets. In 
each test, the planetary models were clearly superior. The modification
that most significantly reduced the $\chi^2$ difference between the
planetary and stellar binary models was to remove the MOA observation
at $t = 4245.10$ or $t = 4245.27$. These modifications allowed 
substantially different stellar binary models that reduced the
$\chi^2$ difference to $\Delta \chi^2 = 40.3$ in each case. Thus,
with the most significant data point removed, the stellar
binary model is still excluded by $6.3\sigma$.

\section{Model Parameter Uncertainties}
\label{sec-param_un}

The microlensing model parameters can be divided up into
three different categories:
\begin{enumerate}
\item Parameters that depend primarily on the planetary deviation
         that are also important for the physical interpretation of the
         lensing event: the mass ratio, $q$, the separation, $d$, and
         the source radius crossing time, $t_\ast$. 
\item Parameters that depend primarily on the overall light curve 
          shape that are important for the physical interpretation of the
          event. These include, the Einstein radius crossing time, $t_E$,
          the source star $I$-magnitude, $I_S$, and the magnitude and 
          angle of the microlensing parallax vector, $\pi_E$ and $\phi_E$.
\item Parameters that do not constrain the important physical parameters
          of the event, such as the time, $t_0$, of the closest approach between
          the source and lens center of mass, the impact parameter, $u_0$, 
          and the angle between the lens axis and source trajectory, $\theta$.
\end{enumerate}

There are a number of different effects that contribute to the uncertainties
in the model parameters, including both discrete and continuous degeneracies.
For MOA-2007-BLG-192, there are four 2-fold degeneracies, with two
in category (2) and two in category (1). The first category (1) degeneracy is the
difference between the cusp approach models and the caustic crossing
models given by Figures~\ref{fig-caustic}(a) and (d), respectively.

The other category (1) degeneracy is the $d  \leftrightarrow 1/d$ degeneracy
\citep{dominik99}, which is a general property of central caustics. It can
sometimes be broken due to the effects of the planetary caustic, which
approaches and attaches itself to the central caustic when $d \rightarrow 1$,
as in the case of Saturn-analog planet OGLE-2006-BLG-109Lc \citep{gaudi-ogle109}.
Alternatively, with reasonably good coverage of the central caustic features
it is possible to break the degeneracy simply by measuring the light curve well,
which was the case with the Jupiter-analog planet in the 
OGLE-2006-BLG-109L system. But, this is much easier to do with relatively
massive planets, and for MOA-2007-BLG-192Lb, this degeneracy is not
broken.

The two 2-fold degeneracies in category (2) parameters are both related
to microlensing parallax \citep{refsdal-par,gould-par1,macho-par1}. 
Microlensing parallax can be described by the
two-dimensional projected Einstein radius vector,  ${\bf \rep}$, which
has an amplitude $\rep = R_E D_S/(D_S-D_L)$, where $R_E$ is the
Einstein radius of the lens system, and $D_L$ and $D_S$ are the lens and 
source distances, respectively. The direction of the ${\bf \rep}$ vector is the
same as the direction of the lens-source relative proper motion. However,
it is generally more convenient to work with the microlensing parallax
vector, $\piEbold \parallel {\bf \rep}$, which has an amplitude equal to
$(1\,{\rm AU})/\rep$.

If a microlensing event was observed by an observer in an inertial reference
frame, there would be no way to determine the direction of lens-source relative
proper motion from the light curve. But for Earth-bound observers, the
acceleration due to the Earth's orbital motion provides a signal of this
direction of motion that can be seen if the light curve measured with
sufficient photometric precision. However, when the direction of lens-source relative
proper motion is pinned down, there is still a reflection symmetry that
remains. If the lens system is reflected about the direction of lens-source
relative proper motion, the resulting light curve will be very similar.
In fact, this is the difference between the orientations of the lens systems
shown in Figures~\ref{fig-caustic}(a) and (c). In each of the panels,
the vertical component of the acceleration of the Earth points downward,
so for the model displayed in panel (c), the acceleration of the Earth is
pushing the system toward closer alignment between the lens and source
(as seen from Earth), while for the 5 other models the acceleration is making
the alignment very slightly worse. 

This reflection transformation takes $u_0 \rightarrow -u_0$, so this is often
referred to as the $u_0 \leftrightarrow -u_0$ symmetry, but in fact, it is only
a good symmetry for high magnification events with $|u_0| \ll 1$. For
a low magnification event with $|u_0| \sim 1$, the $u_0 > 0$ and $u_0 < 0$
solutions will behave differently well after the peak since the acceleration
will keep one solution in better alignment than the other, but for events
like MOA-2007-BLG-192, with $|u_0| \ll 1$, this difference is very small.
The  $u_0 \leftrightarrow -u_0$ degeneracy was first discussed by
\citet{smith_par_acc}.

The second microlensing parallax degeneracy is usually called the
``jerk-parallax" degeneracy \citep{gould-jerkpar} and it is closely
related to a continuous degeneracy discussed by \citet{smith_par_acc}.
To lowest order, the effect of microlensing parallax can be approximated
by assuming the Earth undergoes constant acceleration, and this is
generally a pretty good approximation for events with $t_E \ll 1\,$year.
However, to lowest order in $t_E/1\,$year, we can only 
measure component of $\piEbold$ that is parallel to the Earth's
acceleration. In practice, for most events with a significant microlensing
parallax signal, this means that one component of $\piEbold$ is measured
with significantly greater precision than the other component.
\citet{gould-jerkpar} showed that when expanded to the next order,
the continuous symmetry was removed, but a discrete degeneracy
remained, which was referred to as the jerk-parallax degeneracy
because the next order approximation includes the time derivative 
of the acceleration, or jerk. \citet{multi-par} then showed that for
Galactic bulge sources, a version of the jerk parallax degeneracy persists even
for events with $t_E \simgt 1\,$year because the Galactic bulge is
very close to the ecliptic plane. 
These microlensing parallax degeneracies are broken by the terrestrial
parallax effect \citep{holz_wald}, but for MOA-2007-BLG-192 this effect
is detected at the $\Delta\chi^2 \simlt 1$ level, so these degeneracies
are slightly modified, but not broken.

Because of the four 2-fold degeneracies that we have discussed, there
are 16 local $\chi^2$ minima with $\chi^2$ values within $\Delta\chi^2 < 3$ 
of the best fit solution. These solutions are presented in Table~\ref{tab-fitpar}.
The lowest $\chi^2$ is found for the caustic crossing solutions I and J, but
the source radius crossing time for these solutions, $t_\ast \simeq 0.115\,$days
is quite long given the source star radius of 
$\theta_\ast = 0.50\pm 0.10\,\mu$as estimated below in \S~\ref{sec-lens_prop}.
Using the methods described in \S~\ref{sec-lens_prop}, we find that 
$t_\ast \simeq 0.115\,$days is 1.5-2 times less likely than the 
$t_\ast \simeq 0.065\,$day value favored by the cusp approach models
(\ie\ models A and B). (Without the best fit microlensing parallax constraint,
it is 2 times less likely, and with the constraint it is 1.5 times less likely.) Thus,
the prior constraint on these $t_\ast$ values is equivalent to adding
$\Delta\chi^2 \geq 2\ln 1.5 = 0.81$ (or more)
to the $\chi^2$ of the caustic crossing
models. In addition, these caustic crossing models also require the
star-planet separation to be very close to the Einstein ring radius, which can
be considered to further reduce the {\it a priori} probability of these
caustic crossing models.
So, it is sensible to consider models A and B to be the ``best"
models despite their slightly higher $\chi^2$. In the final analysis, however,
models in the vicinity of all 16 of the models shown in Table~\ref{tab-fitpar}
will be included with their proper statistical weight.

Since these three discrete degeneracies are well understood, we can be confident
that they do not have a significant influence on the behavior of the other
model parameters. Therefore, we can investigate the other possible 
parameter degeneracies by investigating in detail only the $u_0 < 0$, 
$d < 1$, $\pi_{E,N} < 0$ region of parameter space, which contains
models A and I from Table~\ref{tab-fitpar}.

\begin{figure}
\plotone{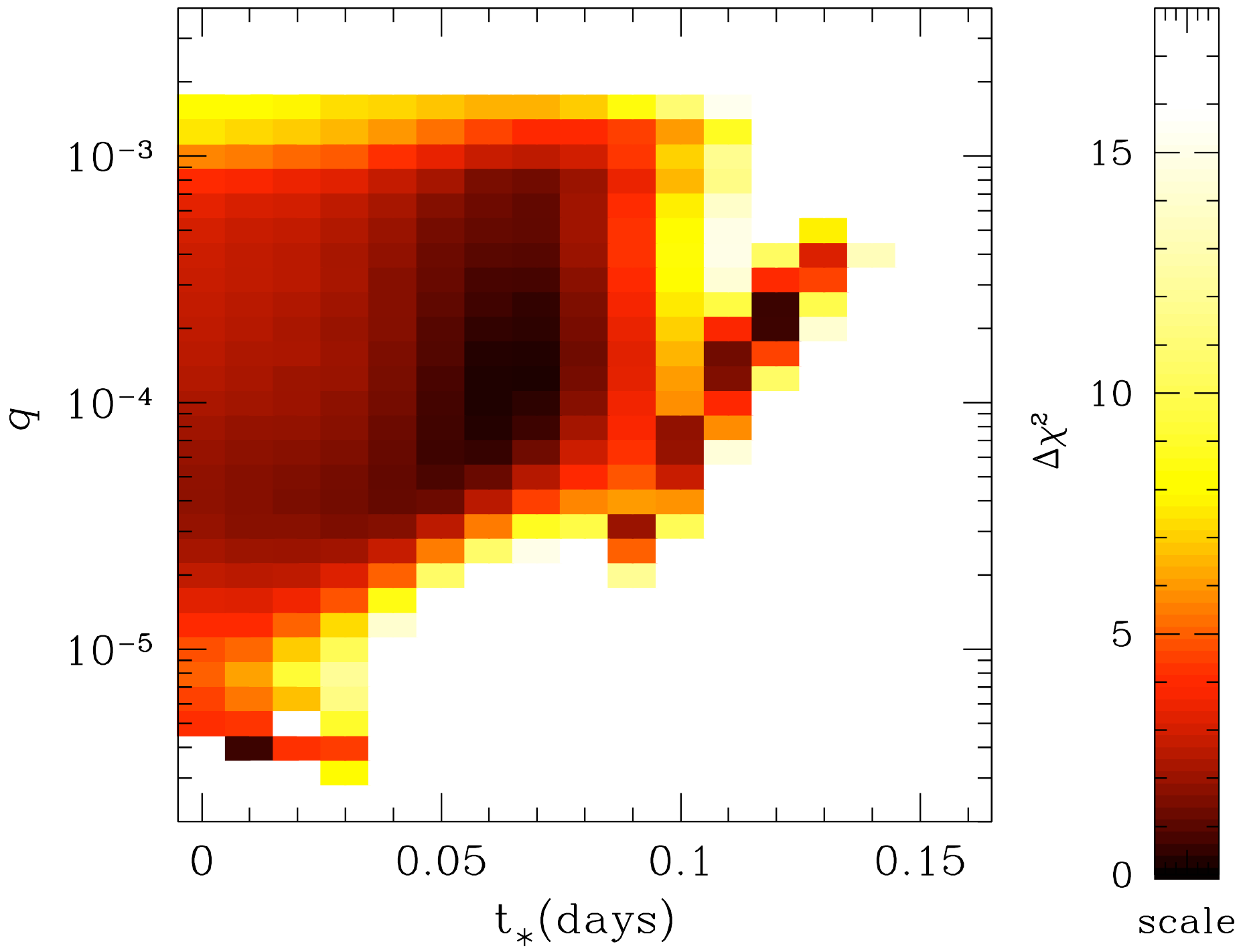}
\caption{The $\chi^2$ difference, or $\Delta\chi^2$, from model A (as
listed in Table~\ref{tab-fitpar})
is plotted for the best fit at each value of the mass ratio, $q$, and the source radius
crossing time, $t_\ast$. All models have $u_0 < 0$, $d < 1$ and
$\pi_{E,N} < 0$.
\label{fig-Ts_eps_grid}}
\end{figure}

Figure~\ref{fig-Ts_eps_grid} shows the results of a search for the best
models on a grid of source radius crossing time, $t_\ast$, and mass
ratio parameters. The three minima represented by the models of
Figures~\ref{fig-caustic}(a), (d), and (e) are apparent. 
Figure~\ref{fig-caustic}(a) is a cusp approach model
that is nearly identical to model 
A of Table~\ref{tab-fitpar}, which has $t_\ast = 0.067$ and 
$q = 1.5\times 10^{-4}$. Model I of Table~\ref{tab-fitpar} is the caustic
crossing model
plotted in Figure~\ref{fig-caustic}(d) with $t_\ast = 0.117$ and 
$q = 2.1\times 10^{-4}$. The final minimum corresponds to the
model of Figure~\ref{fig-caustic}(e), which has  $t_\ast = 0.0066$ and 
$q = 3.9\times 10^{-6}$. The $\chi^2$ for this model, $\chi^2 = 1115.07$,
is actually slightly better than the best models in Table~\ref{tab-fitpar}.

We regard this very low mass planet, quadruple caustic crossing model
as unphysical. As mentioned above, it is {\it a priori} very unlikely that the
last 4 MOA observations on day 4245 managed to hit 3 different
caustic crossings and for the last two MOA observations to have nearly
the same brightness as the earlier two even though most of the light curve
in this region has either much higher or much lower magnification. A much
more likely explanation for the good $\chi^2$ for this model is due to the fact
that its $t_\ast$ value is much smaller than the $\delta t \approx 0.04\,$day interval
between the MOA observations. This
allows the 4 planetary parameters, $d$, $q$,
$\theta$ and $t_\ast$ to be adjusted to fit the 4 MOA observations that are
strongly affected by the planetary deviation. With a larger $t_\ast$ value, the
$0.04\,$day sampling interval provides critical or better sampling of the
planetary features, and the parameters can no longer be adjusted to
account for each of these data points separately. Also, as we shall see
below in \S~\ref{sec-mass}, the lens masses can be determined from
measurements of $t_\ast$ and $\pi_E$, and the small $t_\ast$ provided 
by these models will force a relatively large lens star mass that
must be quite nearby ($\simlt 300\,$pc). Such a lens star would be
much brighter than the upper limit on the lens star brightness if it
is on the main sequence. In fact, the lens star would probably even
be too bright if it were a white dwarf, so these models are also unlikely
from lens star brightness considerations. In \S~\ref{sec-lens_prop},
we will introduce a lens brightness constraint to the modeling
(assuming a main sequence source), and
when applied to these quadruple caustic crossing models, we
find that the best fit has $t_\ast = 0.030\,$days, $q = 4.1\times 10^{-6}$,
and $\chi^2 = 1119.84$. So, it is excluded by just over 2-$\sigma$,
and it also has the low {\it a priori} probability of the quadruple
caustic crossing models against it. As a result, we will classify these
models as unphysical, and not include them in our final statistical
analysis. If this conclusion were wrong, then this model would imply
a planetary host star that is close to the Hydrogen burning limit at
$M \approx 0.08\,\msun$, and the planetary mass would be close to
that of Mars at $\sim 0.1\,\mearth$.

It is apparent from Figure~\ref{fig-Ts_eps_grid} that the grid is not
sampled well enough to map out the detailed
$\chi^2$ behavior for caustic 
crossing and quadruple caustic crossing models. However, we have
already argued that the quadruple caustic crossing models are 
unphysical, and that when we apply the lens brightness constraint,
the best model in this vicinity will be formally excluded by 2-$\sigma$,
so there is little reason to probe these models any further. 
It is also not necessary to sample the caustic crossing models
centered at $t_\ast = 0.117$, $q = 2.1\times 10^{-4}$ at a higher
density, because this region in parameter space does not have
an irregular shape that makes it difficult to sample in a Markov
Chain Monte Carlo. This has been a difficulty with the cusp approach
solutions, and so we have ensured that the sampling of the
grid is dense enough to map out these solutions.

The black regions in Figure~\ref{fig-Ts_eps_grid} essentially
map out the $\Delta\chi^2 \leq 1$ contours. These extend from
$q = 5\times 10^{-5}$ to $q = 4\times 10^{-4}$ and from
$t_\ast = 0.045$ to $t_\ast = 0.075$. These might be considered
the 1-$\sigma$ uncertainty ranges for these parameters. However,
if we consider the $\Delta\chi^2 \leq 4$ contours, we find that the
$t_\ast$ uncertainty is larger than one would predict based on this
1-$\sigma$ range. The red shading in Figure~\ref{fig-Ts_eps_grid} 
maps out the  $\Delta\chi^2 \leq 4$ contours, and this extends from
$t_\ast = 0$ to $t_\ast = 0.092$. So, there is no 2-$\sigma$ lower limit on
$t_\ast$ although the 2-$\sigma$ upper limit on $t_\ast$ is relatively
strong. This is easily understood by considering 
Figures~\ref{fig-caustic}(a)-(c). When the source crossing time is small,
(as in panel (b)) the source passes many source diameters from the 
cusps and the effect of $t_\ast$ on the light curve shape disappears.
This has important implications for the interpretation of the light curve,
because the inferred mass of the lens system scales as $1/t_\ast$, but as 
we shall see in \S~\ref{sec-lens_prop}, the combination of the
microlensing parallax signal and our upper limit on the brightness
of the lens will still allow us to conclude that the primary lens mass
is likely to be substellar.

The $\Delta\chi^2 \leq 4$ contour for the cusp approach solutions 
extends from $q = 1.7\times 10^{-5}$ to $q = 1.08\times 10^{-3}$
and the $\Delta\chi^2 \leq 9$ contour extends from 
$q = 1.2\times 10^{-5}$ to $1.7\times 10^{-3}$. So, the 2-$\sigma$
limits on the cusp approach solutions span a range of 64 in the
mass ratio and the 3-$\sigma$ limits span a range of 140.
The uncertainties in the source radius crossing time, $t_\ast$,
for the cusp approach solution are small at 1-$\sigma$, 
$t_\ast = 0.064 {+0.013\atop -0.019}\,$days, but the 
2-$\sigma$ uncertainty range is quite large, extending
from $t_\ast = 0.093$ down to $t_\ast = 0$. This uncertainty at
small $t_\ast$ is quite important because the angular
Einstein radius and the lens star mass both scale as $1/t_\ast$.
(This is the lens star mass as determined from the combination
of finite source effects and microlensing parallax.)
Fortunately, as discussed below, we do have a lower
limit of $t_\ast \simgt 0.03\,$days from our upper limit on the 
brightness of the lens star, and this will allow us to constrain
the lens star mass in a relatively narrow range.

It is clear from Figure~\ref{fig-Ts_eps_grid} that 
there are acceptable solutions with $t_\ast > 0.093$, which
is the 2-$\sigma$ upper limit for the cusp approach solutions.
These are simply the caustic crossing solutions centered at
the solution shown in Figure~\ref{fig-caustic}(d), which has
$t_\ast = 0.117\,$days. But, aside from the factor of 1.8
difference in the $t_\ast$ values, these caustic crossing 
solutions do not have important differences in any of the
other event parameters. The range of acceptable mass ratio,
$q$, values is very similar to the cusp approach solutions, except
that these solutions do not extend to quite so large $q$ values.

\begin{figure}
\plotone{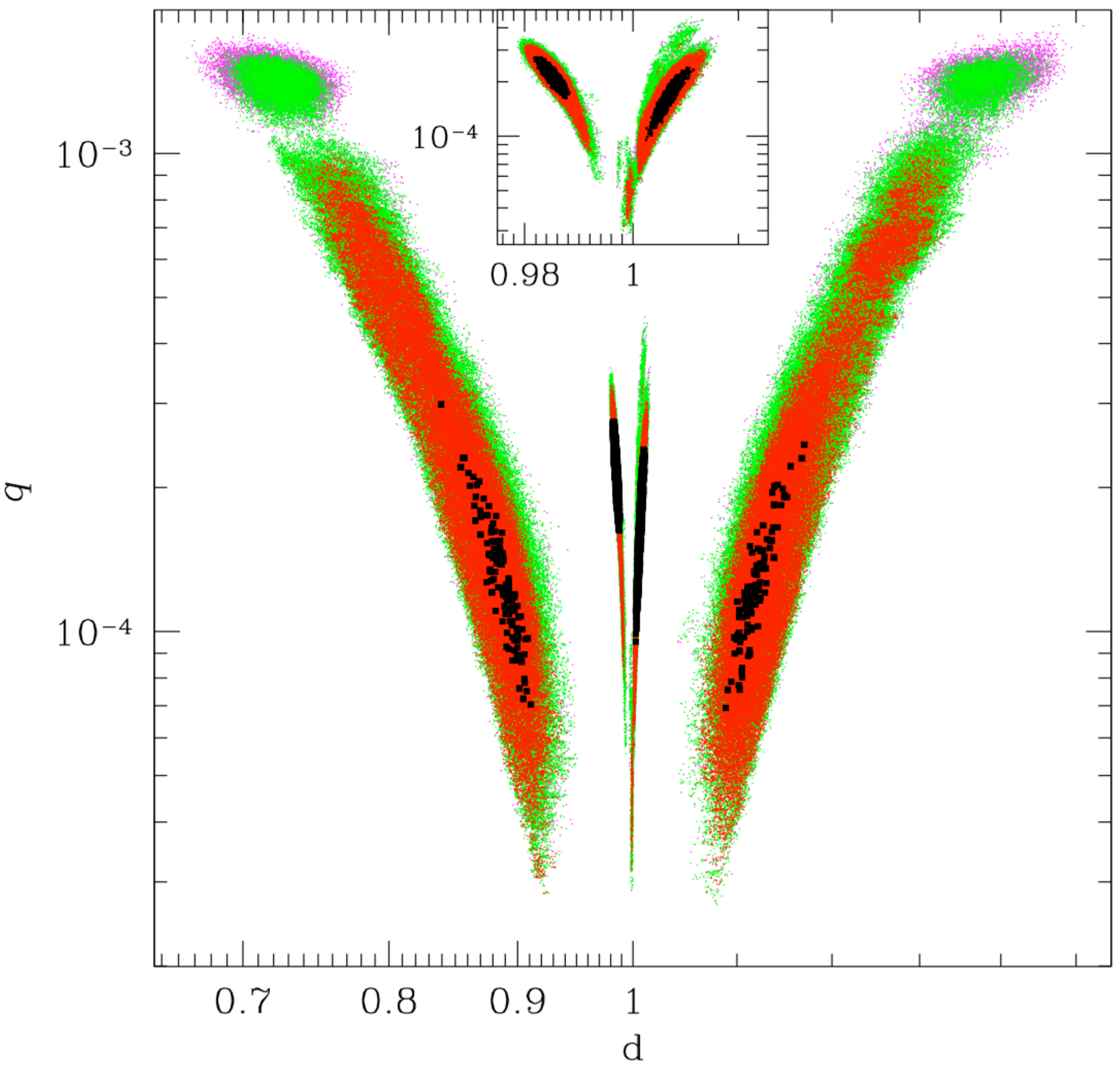}
\caption{The distribution of the planetary mass ratio, $q$ and star-planet
separation, $d$ is plotted for the combined output of 24 Markov Chain Monte Carlo runs,
based on the 16 local $\chi^2$ minima represented by the
parameter sets listed in Table~\ref{tab-fitpar} plus 8 additional runs
to sample the $q > 10^{-3}$ region that was not sampled in the other runs. 
The points are color coded. MCMC links (or light curve models)
within $\Delta\chi^2 \leq 1$, 4, 9, 16, and 25 are plotted as black, 
red, green, magenta, and yellow points. The small inset figure is
just the region of the caustic crossing solutions plotted with a 
greatly expanded x-axis. The points are plotted in
order of decreasing $\Delta\chi^2$, and the yellow and magenta points
are largely covered up by red and green points plotted on top. There
are a total of 809,342 MCMC models plotted in this figure.
\label{fig-mcmc}}
\end{figure}

A more general way to explore the parameter uncertainties is
with a Markov Chain Monte Carlo (MCMC) using an adaptive 
step size Gaussian sampler \citep{cmbeasy}. This and
related methods are
probably the only practical way to explore the parameter
uncertainties for a complicated many dimensional parameter
space, including all the parameter correlations.
Figure~\ref{fig-mcmc} shows a plot of the planetary mass ratio
vs. separation distribution for 
24 MCMC runs. These include 16 runs in the regions of the
16 local minima listed in Table~\ref{tab-fitpar}, plus 8 additional
cusp approach runs designed to cover the area of parameter
space with $q > 10^{-3}$ that was not sampled in the 
other MCMC runs. These areas of parameter space were
not sampled in the runs centered on the appropriate local
minima because of technical difficulties with MCMC
sampling that are discussed below.

Figure~\ref{fig-mcmc}
displays 4 distinct families of solutions. These are separated by
the $d \leftrightarrow 1/d$ degeneracy, and the cusp approach
vs. caustic crossing degeneracy that is specific to this event because
of the incomplete light curve coverage. The two degeneracies
associated with the microlensing parallax effect, the
$u_0  \leftrightarrow -u_0$ and jerk parallax degeneracies,
do not significantly affect the $q$ and $d$ parameters, so
these different models are not separated in Figure~\ref{fig-mcmc}.

The color coding of the points in Figure~\ref{fig-mcmc} indicates 
the $\chi^2$ corresponding to each point. We use $\chi^2 = 1115.46$
as the fiducial that the $\Delta\chi^2$ values are calculated with respect
to, and the points within $\Delta\chi^2 \leq 1$, 4, 9, 16, and 25 are 
plotted in black, red, green, magenta, and yellow. The magenta and yellow
points are generally covered up by the red and green points. Note that this
color coding is for display purposes only. Points of different colors are
not distinguished in subsequent calculations. 

It appears from Figure~\ref{fig-mcmc} that there might be separate
local $\chi^2$ minima at $q\approx 1.3\times 10^{-3}$ and 
$d \approx 0.73$, 1.37. In fact, these are the locations of the 8 additional,
out of equilibrium, runs that were done to sample this region parameter
space that was identified in the grid search (see Figure~\ref{fig-Ts_eps_grid}),
but were not included in the MCMC runs centered on the local minima of
Table~\ref{tab-fitpar}. These are simply areas of parameter space that
are difficult to sample with the equilibrium MCMC runs, and they are
not separate minima.

One notable feature of Figure~\ref{fig-mcmc}  is that there are relatively
few points with $\Delta\chi^2 \leq 1$ (even though the $\Delta\chi^2 \leq 1$ points
are plotted with larger dots than the other points). Of course, with 9 fit
parameters, we expect $\sim 110$ times as many $\Delta\chi^2 \leq 1$
points as $\Delta\chi^2 \leq 4$ points from the statistics of Gaussian random
variables. But, in Figure~\ref{fig-mcmc} the ratio of $\Delta\chi^2 \leq 4$
points to  $\Delta\chi^2 \leq 1$ points is about 2500.
There are two reasons for this.
First, Figure~\ref{fig-mcmc} contains 24 different Markov chains: 16 centered
on the local minima solutions given in Table~\ref{tab-fitpar}, and 8 more
runs for the cusp approach $q\approx 1.3\times 10^{-3}$ regions.
15 of these 24 regions have a minimum
$\Delta\chi^2 > 1$ and 3 local minima have $\Delta\chi^2 \sim 1$,
and so these branches will produce no or very few MCMC points with
$\Delta\chi^2 \leq 1$. Second, as we remarked in the discussion of 
Figure~\ref{fig-Ts_eps_grid}, the $\Delta\chi^2 \leq 1$ region for each 
of the cusp approach solutions is quite small compared to the 
$\Delta\chi^2 \leq 4$ region. That is, the shape of the $\chi^2$
surfaces for each of the cusp approach solutions is far from the 
multidimensional parabolic shape that is typically expected.
The caustic crossing solutions do have $\chi^2$ surface shape
that is closer to parabolic, and therefore, they have the 
more typical distribution of $\Delta\chi^2 \leq 1$, 4, and 9 points,
although only 4 of these solutions a a minimum $\Delta\chi^2 \leq 1$.
 
An additional feature of the cusp approach solutions that is apparent
in Figure~\ref{fig-mcmc} is that the $\Delta\chi^2 \leq 1$ points
lie at $d$ closer to 1 than the $\Delta\chi^2 \leq 4$ and 9 points.
This is a manifestation of the highly asymmetric 2-$\sigma$ error bars
for $t_\ast$ that we encountered in Figure~\ref{fig-Ts_eps_grid}.
The small $t_\ast$ solutions also have $d$ further from 1 than the 
large $t_\ast$ solutions, so this same feature can be seen in both
these figures.
In fact, this effect would be even more pronounced in Figure~\ref{fig-mcmc}, 
but we have excluded most of the $t_\ast \leq 0.03\,$days parameter space
by imposing a constraint on the lens star brightness. We will discuss
the physics behind this constraint in \S~\ref{sec-lens_prop}, but part
of the motivation for including this constraint at this stage of the analysis is
computational.
 
As we have seen in the discussion of Figure~\ref{fig-Ts_eps_grid}, the
$\chi^2$ surface for models of this event is rather complicated, and this
can make it difficult to sample the full parameter space of allowed models.
In order to sample parameter space more efficiently, we have
implemented an an adaptive step-size Gaussian sampler following
\citep{cmbeasy}. We calculate the
correlation function of the first 1000 steps in each chain.
New models parameters to be tested for inclusion in the chain are 
selected using the linear combination of parameters that diagonalizes the
correlation function. This procedure usually allows relatively large steps
through parameter space, while still maintaining a large probability that
the new model will have a $\chi^2$ low enough to be included in the 
chain. For MOA-2007-BLG-192, this improvement is not as dramatic as with
other events because the shape of the $\chi^2$ near the minima tends
to depend on the parameters in non-linear ways that are therefore not
captured by the correlation function. Sometimes, this can be cured
or reduced by changing
variables. For example, we have found that a Markov Chain using
$\log (q)$ as a parameter instead of $q$ provides a much more rapid
sampling of parameter space.

\begin{figure}
\plotone{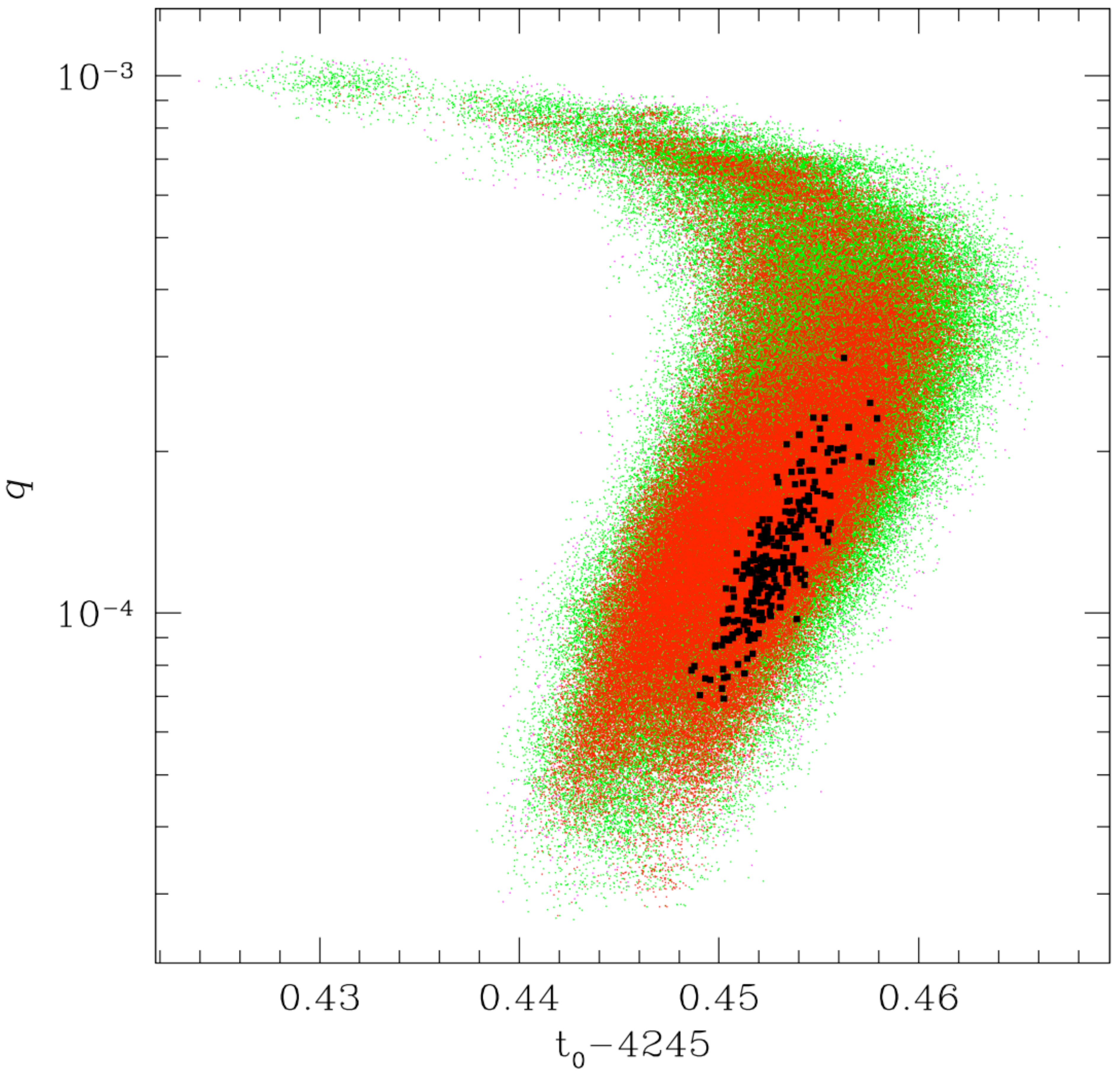}
\caption{The distribution of the planetary mass ratio, $q$, and the
time of closest alignment, $t_0$ are plotted for the
combined output of 8 Markov Chain Monte Carlo runs,
based on the cusp approach models A-H of Table~\ref{tab-fitpar}.
The color coding is the same as in Figure~\ref{fig-mcmc}
\label{fig-mcmc_qt0}}
\end{figure}

The use of $\log (q)$ as a parameter  does not completely cure this
problem due to the shape of the $\chi^2$ surface. We can see this
in Figure~\ref{fig-mcmc_qt0}, which shows the distribution
of the $q$ and $t_0$ parameters for the 8 cusp approach $\chi^2$
minima listed in Table~\ref{tab-fitpar}. The distribution is approximately
described by an ellipse for $q < 5\times 10^{-4}$, but for 
$q \simgt  5\times 10^{-4}$ the distribution veers off at about a $90^\circ$
angle from the direction of the ellipse. This means that for 
$q \simgt  5\times 10^{-4}$, the adaptive step-size Gaussian sampler
is not efficient at selecting parameters for new links in the chain, and 
a much higher fraction of parameter sets are rejected and don't end
up in the Markov chain. Furthermore, this also causes the correlation
length of the Markov Chain to grow, which decreases the efficiency of
the sampling even further. The effect of this longer correlation length
can be seen in both Figure~\ref{fig-mcmc} and Figure~\ref{fig-mcmc_qt0}.
For $q \simgt  5\times 10^{-4}$, the distribution of points is clustered
into stripes that run nearly horizontally in Figure~\ref{fig-mcmc_qt0}
and are tilted somewhat in Figure~\ref{fig-mcmc}. As a result of this
effect, the statistical noise in the Markov Chain results is significantly
larger for $q \simgt  5\times 10^{-4}$ than for smaller values of $q$.

This MCMC sampling difficulty is likely to be responsible for the 
poor sampling of the $q > 10^{-3}$ models in these cusp approach
Markov chains. In fact, most of the $q > 10^{-3}$ models in the 
additional runs we have done have $t_0 - 4245 < 0.42$, so they 
would be off the left hand side of these plots.

The most serious problem we encountered with Markov Chain 
sampling involved the
parameter $t_\ast$. Some chains would run very long 
with $t_\ast \simgt 0.03\,$days, whereas other chains would run very long with
$t_\ast < 0.01\,$days, and it was very rare for each chain to cross from
one region to the other. Thus, it was difficult to get well sampled Markov
chains without a very large number of steps. However, as
we shall see in \S~\ref{sec-lens_prop}, the models with 
$t_\ast \simlt 0.02\,$days are largely excluded by our constraints
on the source star brightness. Therefore, an effort to fully explore this
region of parameter space would be of little interest, and so we
have applied this constraint to our MCMC calculations.

\section{Lens System Characterization}
\label{sec-lens_prop}

\subsection{Lens System Mass Determination}
\label{sec-mass}

The primary difficulty in the interpretation of most microlensing events
is the fact that the lens mass, distance, and velocity only affect a single
measurable event parameter, the Einstein radius crossing time, $t_E$.
But, the situation is usually improved for planetary microlensing events
because most planetary microlensing events have intrinsic features of very 
short duration that allow the source radius crossing time to be measured. Since the
source star angular radius, $\theta_\ast$ is usually known from the source 
brightness and color, the measurement of finite source size effects generally
allows the angular Einstein radius, $\theta_E = \theta_\ast t_E/t_\ast$, to
be determined. The measurement of $\theta_E$ reduces the lens system
uncertainty to a single parameter family of solutions, so that the lens system
mass and relative velocity will be determined as a function of the distance
to the lens, $D_L$.

The remaining lens system uncertainty can be removed when the
microlensing parallax effect is measured. Microlensing parallax can
be described \citep{exoplanet_book} by the projected Einstein radius, 
$\repbold$, which has a magnitude, $\rep$, equal to the Einstein radius 
projected (from the source) to the position of the observer and a direction
parallel to the lens-source relative proper motion. The magnitude of the
projected Einstein radius, $\rep$, can then be directly related to $\theta_E$
and the lens system mass. In the small angle approximation, the deflection
angle for a lens system with perfect alignment is
\begin{equation}
\alpha = {\rep\over D_L} = {4GM\over \theta_E D_L c^2} \ ,
\label{eq-angle_m}
\end{equation}
where the first expression for $\alpha$ comes from geometry, and the second
expression for $\alpha$ is just the general relativistic formula for light 
deflection by a point mass. We can solve equation~(\ref{eq-angle_m})
for $M$ to yield
\begin{equation}
M = {\theta_E \rep c^2\over 4G}  \ ,
\label{eq-mass}
\end{equation}
for the lens system mass. This method has been used to
determine the mass of a few lens systems
 \citep{planet-er2000b5}, including the Jupiter/Saturn analog
 system OGLE-2006-BLG-109L \citep{gaudi-ogle109,bennett-ogle109}.
 But the situation for MOA-2007-BLG-192L is somewhat more complicated
 because $t_\ast$, $\rep$, and even $\theta_\ast$ are not perfectly 
 measured. So, we must factor all of these uncertainties into our
 estimate of the lens mass.

\subsection{Source Star Angular Radius}
\label{sec-src_rad}

The source star radius is normally determined \citep{yoo_rad} from its
brightness and color using the empirical color-radius relations of
\citet{kervella_dwarf}. However, in this case, we have no measurement
of the source star color because the event was not realized to be a planetary
event until after the magnification had dropped significantly. Nevertheless,
it is still possible to estimate the source star color, because stars of similar
magnitude in the direction of the Galactic bulge
are observed to have a relatively narrow range of colors
\citep{holtz}, due to the fact that most of the stars of this brightness
($I\simeq 21.45$) seen toward the bulge are
actually Galactic bulge main sequence stars. Thus, we can estimate the
color of the source based upon its brightness and upon the observed
colors of stars of similar intrinsic brightness.

Before we can compare to the \citet{holtz} HST observations of Baade's Window,
we must adjust for the difference in extinction and distance between the 
field of MOA-2007-BLG-192 and Baade's Window. This is most easily done
by comparing locations of the centroid of the red clump giant feature of the
color magnitude diagrams centered upon the  \citet{holtz} Baade's Window
and MOA-2007-BLG-192. From the calibrated OGLE-II database, we find
\begin{eqnarray}
(I,V-I)_{\rm clump,MOA192} = &(15.74, 2.16)   \   , \ \ \ \   \\
(I,V-I)_{\rm clump,Holtz} = &(15.15,1.62) \ \   {\rm and}   \\
\left[\Delta I, \Delta (V-I)\right]_{\rm clump} = &(0.59, 0.54) \ , \ \ \ \ \ \ \
\end{eqnarray}
for the magnitude and color offset between the MOA-2007-BLG-192 field
and the Baade's Window field of \citet{holtz}. Thus, if the MOA-2007-BLG-192
source star was moved to Baade's Window, we would expect its apparent
magnitude to change from  $I_s = 21.45$ to $I_{s{\rm BW}} = 20.86$ in
Baade's Window. We then estimate the $V-I$ color
that the source would have from the average of 1206 stars observed by
\citet{holtz} with magnitudes in the range $20.76 \leq I_{s{\rm BW}} \leq 20.96$.
After (iteratively) removing 29 3-$\sigma$ outliers from the \citet{holtz} star list, we find
$(V-I)_{s{\rm BW}} = 1.69 \pm 0.20$, which converts to 
$(V-I)_s = 2.23\pm 0.20$ in the MOA-2007-BLG-192 field. (The 3-$\sigma$ outliers
are almost entirely redder foreground stars that would have a low probability
of being microlensed.)

In order to determine the source star radius, we will need to correct the
source star magnitude and color for extinction. This can be done by comparing
the observed magnitudes of the red clump giant feature 
to the values expected based upon red clump giants observed locally (with 
small theoretical corrections for age and metalicity). From
\citet{clump_agemet1} and \citet{clump_agemet2} we have
$M_I = -0.25\pm 0.05$,  $M_V = 0.79\pm 0.08$ and $V-I = 1.04\pm 0.08$
for the centroid of the Galactic bulge red giant clump. (We have assigned the
error bars for these values based upon the size of the theoretical corrections to
the red clump giant magnitudes.)
Due to the bar-like nature of the bulge, the stars in the field of MOA-2007-BLG-192 
at Galactic coordinates, $\ell = 4.0309^\circ$ and $b = -3.3877^\circ$
are expected to be slightly closer to us than the stars in the Holtzman field. 
From \citet{rat_bar}, we estimate a distance modulus of $DM = 14.38 \pm 0.07$,
assuming a distance of $8.0\,$kpc to the Galactic Center, so we estimate the 
dereddened magnitude and color of the red clump giant centroid to be
$I_{0,\rm clump} = 14.13 \pm 0.09$ and 
$(V-I)_{0,\rm clump} = 1.04 \pm 0.08$. This implies an extinction of
$A_I = 1.61 \pm 0.10$ and reddening of $E(V-I) = 1.12 \pm 0.09$,
which is quite similar to the value from the map of 
\citet{sumi_ext_map}, ($E(V-I) = 1.04$.

These extinction and reddening values imply $R_{VI} \equiv A_V/E(V-I) = 2.44$,
which implies $R_V \equiv A_V/E(B-V) = 3.04$ according to 
the reddening formula of \citet{reddening}. This is quite similar to the ``standard"
value of $R_V = 3.1$, in seeming contradiction to claims of anomalous extinction
towards the Galactic bulge. However, the strongest evidence for this anomalous
extinction \citep{udal_anon_ext} involves the difference between the extinction
along nearby lines-of-sight instead of the average extinction along any single
line-of-sight. But, the difference between nearby lines-of-sight is likely to be
dominated by the dust far from the position of the Sun, so it is more likely to
be anomalous than the average along the line of sight to a bulge field.

With this adopted reddening and extinction values, the 
dereddened source magnitude and color become
$I_{s0} = 19.84 \pm 0.24$ and Ê$(V-I)_{s0} = 1.11 \pm 0.24$. 
\citet{kervella_dwarf} provide a set of relations to estimate the stellar
angular radius from its magnitude and color, but there are two complications
with the $V$-$I$-radius relations. First, these relations use the Johnson-$I$ band
magnitudes, whereas all the other $I$ magnitudes reported in this paper
use the Cousins system. Using the 3 stars (GJ 105 A, GJ 570 A, and
$\epsilon$ Ind A) in the \citet{kervella_dwarf} sample 
with a similar color to our estimate, $(V-I)_{s0} = 1.11 \pm 0.24$, for the source
star, we find $I_{\rm Johnson} = I_{\rm Cousins} + 0.30\pm 0.03$. The second
complication is that the $V$-$I$-radius relations are non-linear. So, we use
a cubic fit to the color and radius values of \citet{kervella_dwarf} to yield
$\theta_\ast = 0.50 \pm 0.10\,\mu$as for the angular source radius for the assumed
source magnitude of $I_s = 21.44$. When this $\theta_\ast$ value is used below
to estimate the lens system mass, the source brightness will be allowed to vary 
somewhat from this assumed value. So,
we will also include the scaling with
source brightness: $\theta_\ast = (0.50\pm 0.10) 10^{0.2(21.44-I_s)}$. 
In principle, we should also include the effect on our estimate of the source
color, but this effect is much smaller than the uncertainty in the color, so we
neglect it.

\subsection{Microlensing Parallax}
\label{sec-par}

Our primary conclusion that MOA-2007-BLG-192 lens primary is likely to
have a sub-stellar mass derives primarily from the microlensing parallax 
\citep{refsdal-par,gould-par1,macho-par1} signal. Of course, 
it is always possible for orbital motion of the source to mimic the
microlensing parallax effect \citep{multi-par}. This is often referred to
as the ``xallarap" effect, and we discuss the possibility that the observed
signal may be due to xallarap rather than parallax in the Appendix
(\S~\ref{sec-xal}). We show that it is unlikely, but not impossible, 
that the apparent microlensing parallax signal is really due to xallarap,
and in \S~\ref{sec-conclude} we discuss future observations that could
rule out the xallarap hypothesis. For the remainder of this section, we
will assume that the observed signal is really due to microlensing
parallax.
It is important to understand the microlensing parallax measurement
in some detail because of their implications for the interpretation of
this event. In \S~\ref{sec-mass}, we found it convenient to use the
projected Einstein radius vector, $\repbold$, to describe the implications of
a microlensing parallax measurement. But this is not such a convenient variable
to use for fitting light curves, because $\rep \rightarrow \infty$ when the 
microlensing parallax signal is weak. Instead, we prefer to work with the
microlensing parallax vector, $\piEbold \parallel \repbold$, which has a 
magnitude, $\pi_E = 1\,{\rm AU}/\rep$. 

\begin{figure}
\plotone{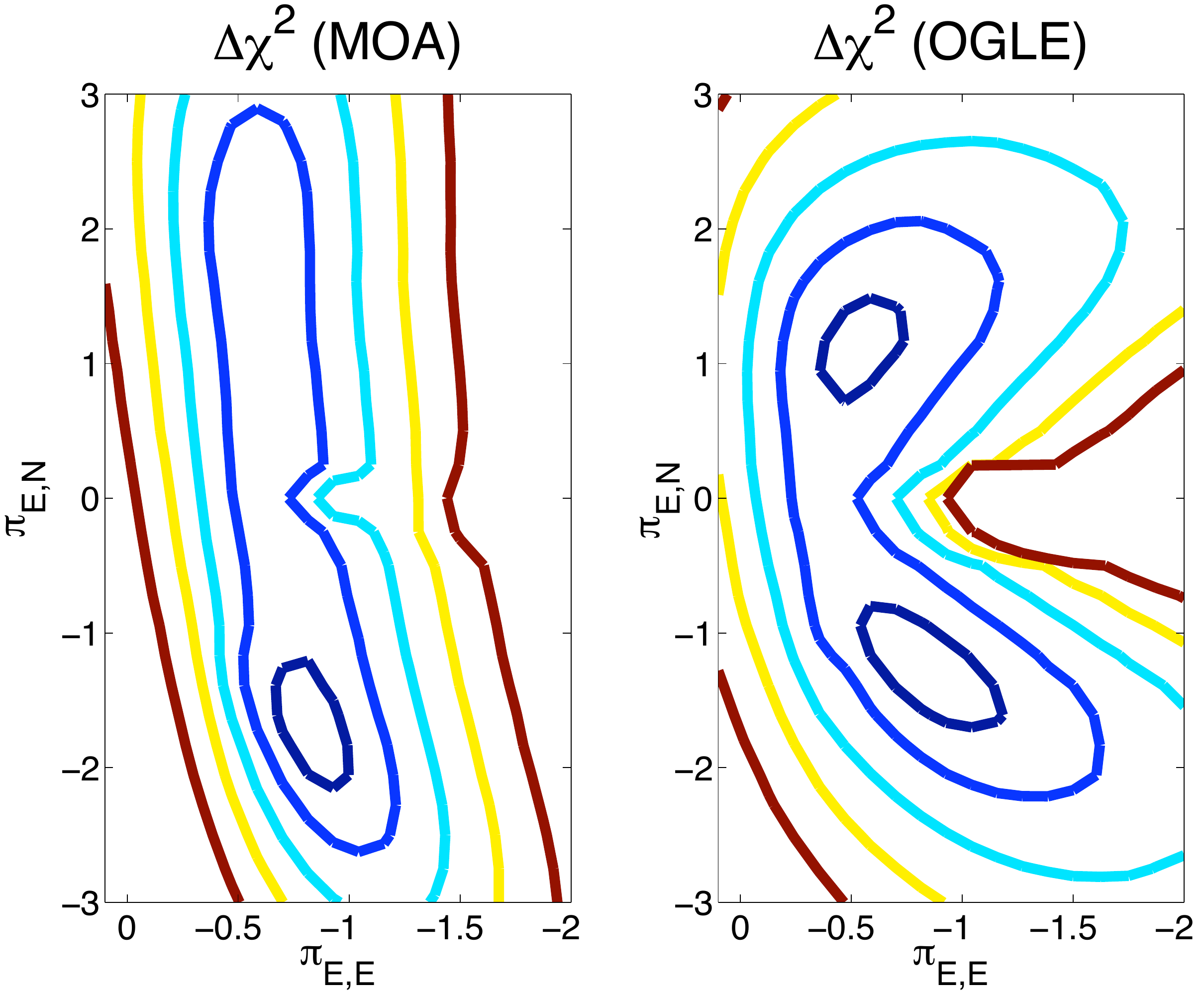}
\caption{$\Delta\chi^2$ contours are plotted for the East vs.
North components of the microlensing parallax vector, $\piEbold$.
These are based on joint fits to the MOA and OGLE data with the
data in the region of the planetary signal ($4244.8 < t <  4246.3$) removed.
The left and right panels indicate similar signals in the MOA and OGLE data.
The contour levels plotted are $\Delta\chi^2 = 1$ (dark blue), 4 (blue), 9 (cyan),
16 (yellow), and 25 (brown).
\label{fig-piE_cont}}
\end{figure}

Figure~\ref{fig-piE_cont} shows the $\Delta\chi^2$ contours for microlensing
parallax fits to the MOA-2007-BLG-192 light curve
with the region of the planetary signal removed. Observations with 
$4244.8 < t <  4246.3$
are excluded, and a single-lens parallax model was fit to the data.
This figure reveals a number of
the degeneracies discussed in \S~\ref{sec-param_un}. From both the MOA and
OGLE data, it is clear that $\piEbold$ is constrained much more tightly in one
direction (nearly the E-W direction) than the other \citep{gould-diskorhalo}. 
Overall, the microlensing
parallax signal is detected more strongly in the MOA data (at $> 5$-$\sigma$)
than in the OGLE data (at $\sim 3$-$\sigma$). This is probably due to the
fact that there are $>7$ times as many MOA observations as OGLE observations
on the magnified part of the light curve. But, the OGLE telescope generally has
better seeing than MOA, and this may account for the slight breaking of the
(continuous) constant acceleration degeneracy \citep{smith_par_acc}
seen in the OGLE data.
The discrete jerk-parallax degeneracy \citep{gould-jerkpar} is also seen
and is broken at a relatively low level of confidence. Note that these same degeneracies
do not apply to the terrestrial parallax effect \citep{holz_wald}, which is due to
the different locations of the observatories on the Earth. This effect can only
be detected in the rapidly varying parts of the light curve, so it is not included
in Figure~\ref{fig-piE_cont}. Similarly, the $u_0 \leftrightarrow -u_0$ degeneracy is 
essentially exact with the light curve peak removed, so we haven't 
considered the $u_0 > 0$ and $u_0 < 0$ solutions separately. However, when
we do consider complete planetary models with parallax, we find that
the terrestrial parallax effect does contribute to the resolution of these
degeneracies, adding $\Delta\chi^2 = 0.4$ to the difference between the 
$\pi_{E,N} < 0$ and $\pi_{E,N} > 0$ solutions. The total $\chi^2$ improvement
from adding microlensing parallax to the best cusp approach solution is
$\Delta\chi^2 = 40.53$.

\begin{figure}
\plotone{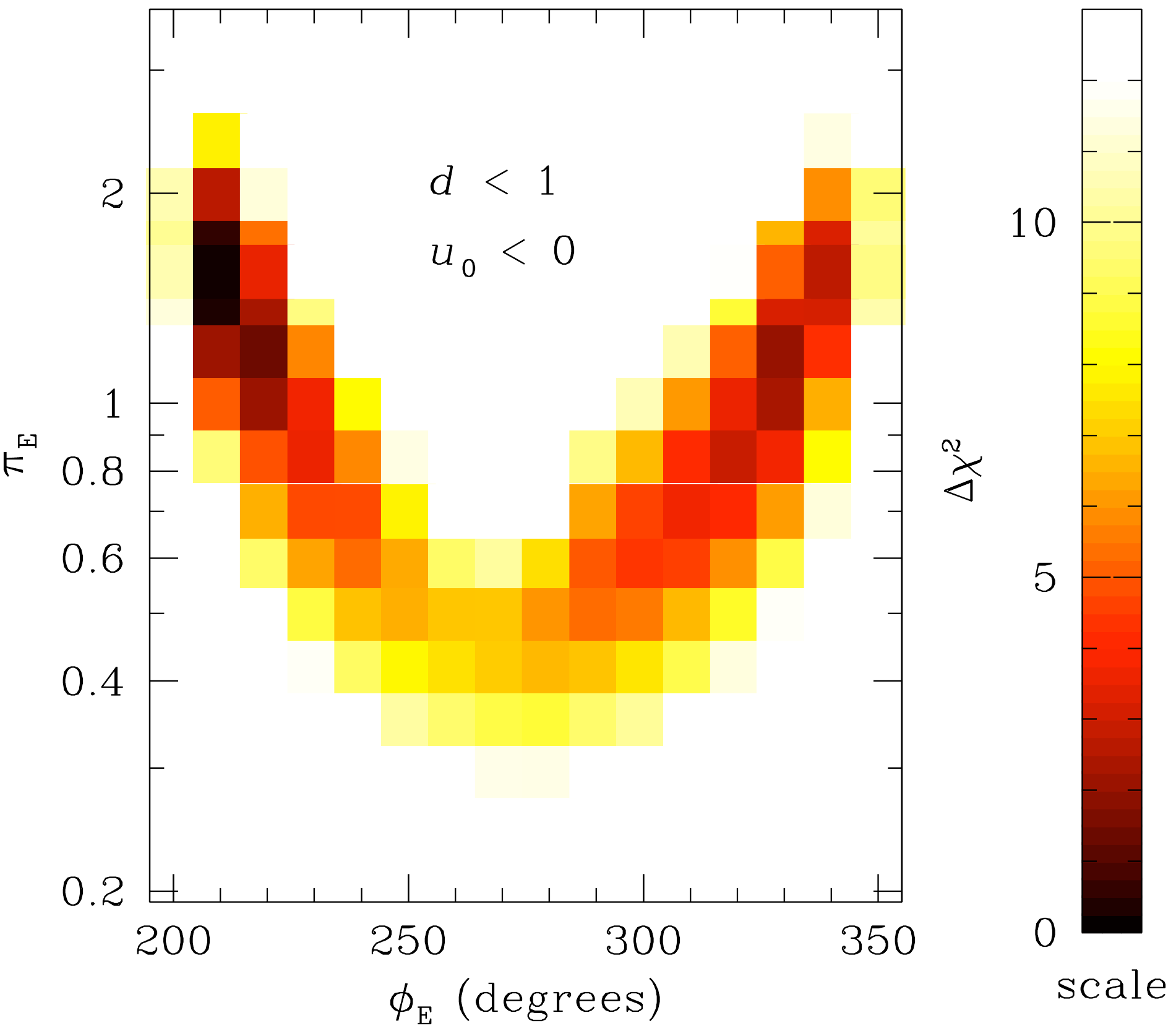}
\caption{The regions of microlensing parallax parameter space that
are consistent with the MOA-2007-BLG-192 light curve are indicated
by the distribution of $\Delta\chi^2$ from the best fit model. 
The $\pi_E$ and $\phi_E$ values that are not shown and those that
are indicated by white squares all give
$\Delta\chi^2 >  12$ larger than the minimum value.
\label{fig-piEgrid}}
\end{figure}

Equation~(\ref{eq-mass}) can be rewritten as
\begin{equation}
M = {\theta_E c^2 {\rm AU}\over 4G\pi_E}  \ ,
\label{eq-mass_pi}
\end{equation}
so it is the magnitude of $\pi_E$ that is directly related to the lens mass.
Thus, it is instructive to plot a $\chi^2$ surface map using polar
coordinates, as in Figure~\ref{fig-piEgrid}. This figure 
shows such a $\chi^2$ surface map using polar
coordinates such that the North component of $\piEbold$ is given by
$\pi_E \cos\phi_E$ and the East component is given by $\pi_E \sin\phi_E$
for the $d < 1$, $u_0 < 0$ branch of the cusp approach solutions. (This is the
region including fits A and E of Table~\ref{tab-fitpar}.) The analogous plots
for the 7 other parameter regions (corresponding to the 14 other models
listed in Table~\ref{tab-fitpar}) are all very similar. 

\begin{figure}
\plottwo{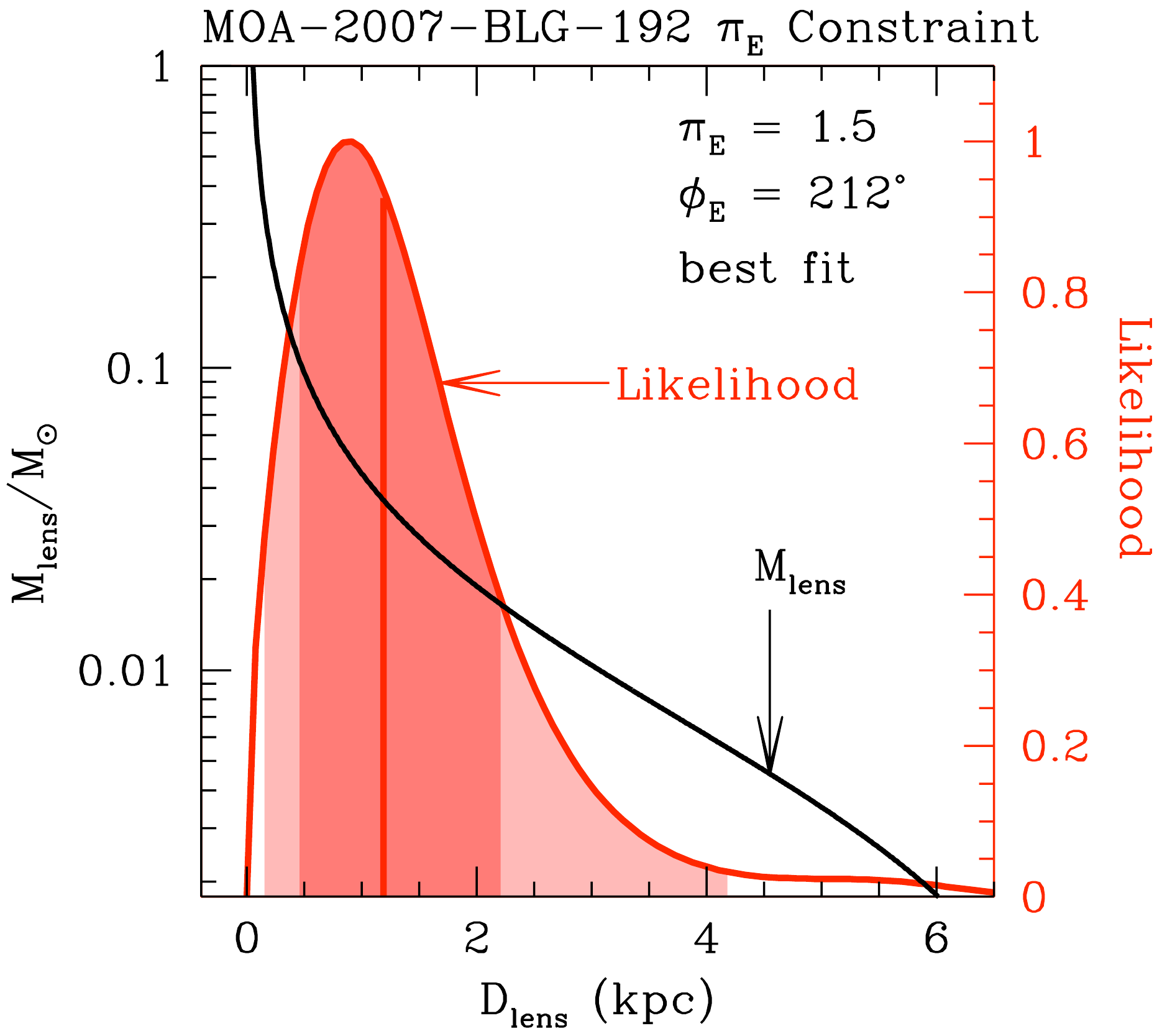}{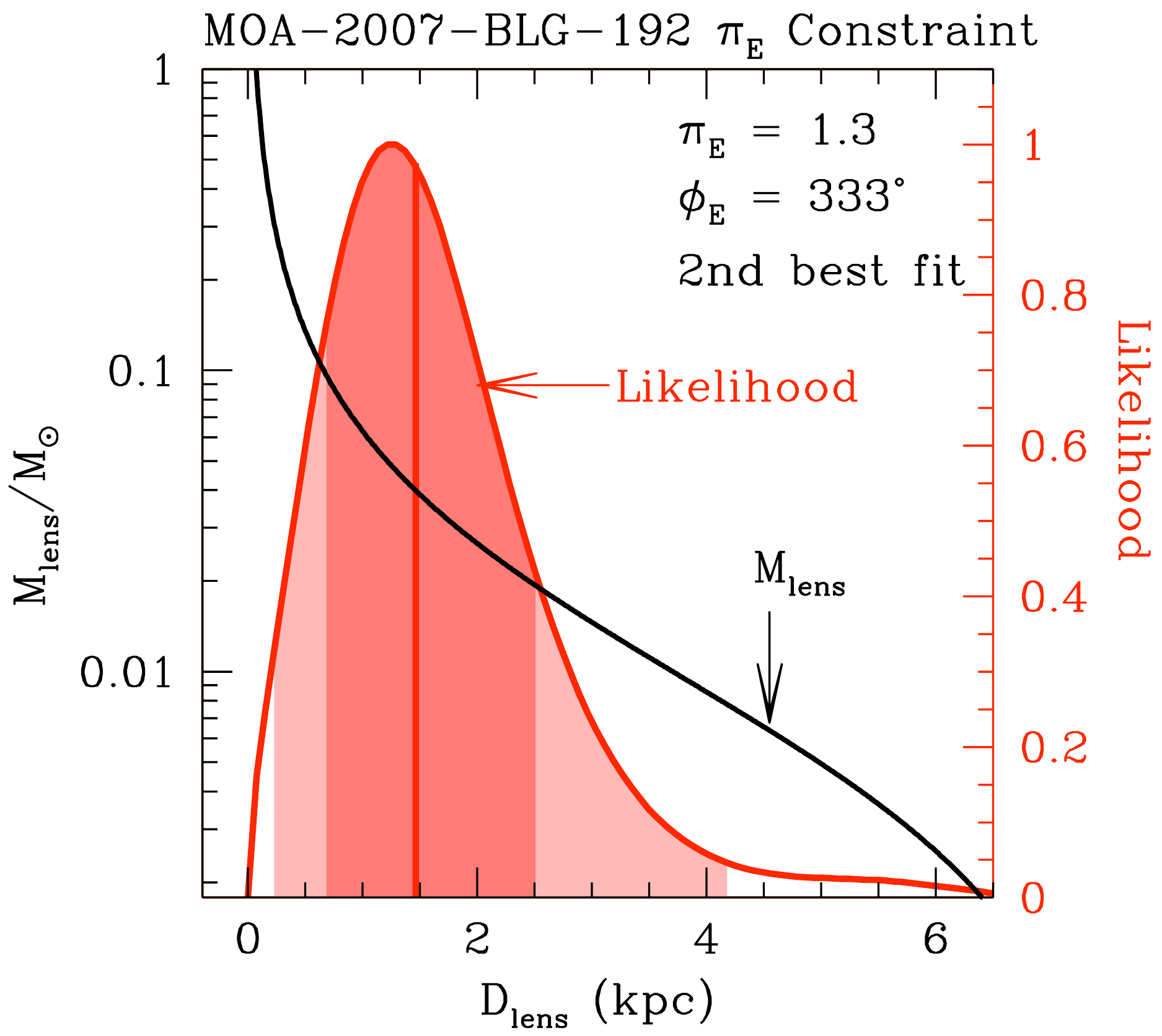}
\caption{
The mass-distance relations are plotted in black
for the two local minima in the
microlensing parallax parameter $\chi^2$ surfaces shown in 
Fig.~\ref{fig-piEgrid}. The red curves show the probability
distributions from a Bayesian analysis that compares the
$\vpbold$ for each model to a standard Galactic model.
The vertical red lines indicate the median distance and lens
primary mass and the light red shaded regions indicate the 1-$\sigma$ and
2-$\sigma$ limits on the lens distance and mass. The median and
1-$\sigma$ limits for the lens star mass are $M = 0.036 {+0.057\atop -0.020} \msun$
and $M = 0.039 {+0.051\atop -0.020} \msun$ for the best and 2nd best fits,
respectively. The 2-$\sigma$ ranges are $0.005\msun \leq M \leq 0.36\msun$ and
$0.007 \leq M \leq 0.31\msun$.
\label{fig-masslike}}
\end{figure}

If we ignore the constraints on $t_\ast$ (and therefore $\theta_E$) for the 
moment, we can use the best fit microlensing parallax values by themselves 
to estimate the lens mass. First, a measurement of $\pi_E$ implies the
following relation, 
\begin{equation}
M = {c^2\over 4G} \left({1\,{\rm AU}\over \pi_E}\right)^2 
   {D_S - D_L\over D_S  D_L} \ ,
\label{eq-m_piE}
\end{equation}
which we can take to be a mass-distance relation, since the source distance,
$D_S$, is known with reasonable accuracy. These curves are plotted
for models A and E as the black curves in the left and right
panels (respectively) of Figure~\ref{fig-masslike}. The "best fit"
plot for model A, is nearly identical to the mass-distance plots for
the other models with the "best fit" parallax parameters
($\pi_E \approx 1.5$, $\phi_E \approx 212^\circ$), namely
models B-D and I-L. Similarly, the "2nd best fit" plot for model E
is nearly identical to the plots for the other $\phi_E \approx 333^\circ$
models (F-H and M-P).

We can make use of our knowledge of Galactic kinematics to constrain
the lens distance, $D_L$, and mass via Equation~(\ref{eq-m_piE}), if we
use yet another parameter 
to describe the microlensing parallax measurement. 
The projected velocity vector, 
\begin{equation}
\vpbold \equiv {\tilde {\bf r}_E\over t_E}
         = {{\rm AU}\over \pi_E^2 t_E}\piEbold \ ,
\label{eq-vp_def}
\end{equation}
depends only on the lens
and source kinematics, so we can use it for a Bayesian analysis
of the source distance following \citet{macho-par1} without 
any need to insert a prior for the lens star mass function. 

This introduces an additional subtlety into the analysis. While the scalars
$\theta_E$, $\pi_E$ and $\rep$ are independent of the reference frame
used, any variables related to timing or the direction of the lens-source
relative motion will depend on the reference frame that is used. Conceptually,
it is easiest to deal with microlensing parallax in the heliocentric reference
frame \citep{gould-par1,macho-par1}. But \citet{gould-jerkpar} pointed out
that the details of the microlensing parallax signals are easier to understand
in a geocentric frame
that is at rest with respect to the Earth at some time close to the peak 
magnification of the microlensing event. We have therefore used the
geocentric frame at rest with respect to the Earth at $t = 4245$
in our parallax analysis. It is certainly possible to 
continue to use this geocentric frame in our comparison to Galactic
models, but it is far more convenient to use the heliocentric reference
frame to compare to Galactic models because the heliocentric value
of $\vpbold$ does not depend on the phase of the Earth's orbit.
Therefore, we convert from geocentric to heliocentric reference
frame via
\begin{equation}
\tilde {\bf v}_{\rm hel} = 
    \tilde {\bf v}_{\rm geo} + {\bf v}_{\oplus,\perp}(t=4245)
\label{eq-heliogeo}
\end{equation}
where ${\bf v}_{\oplus,\perp}(t=4245) = (+1.3,+25.7){\rm km\,s^{-1}}$
(north, east) is the velocity of the Earth projected onto the plane 
of the sky at the peak of the event.

Our Bayesian analysis assumes a double-exponential disk with
parameters based on \citet{reid_mdwarf_kin} and 
\citet{hipp_loc_kin}, and a Galactic bar model from 
\citet{hangould-mpar} with rotation that matches the analysis of
\citet{rat_bar}. A Bayesian analysis with this Galactic model yields
the likelihood functions given by the shaded red curves in 
Figure~\ref{fig-masslike}. The implied lens mass
is almost identical for the ``best fit" and ``2nd best fit" cases,
with the predicted masses given by $M = 0.036 {+0.057\atop -0.020} \msun$
and $M = 0.039 {+0.051\atop -0.020} \msun$, respectively.

\begin{figure}
\plotone{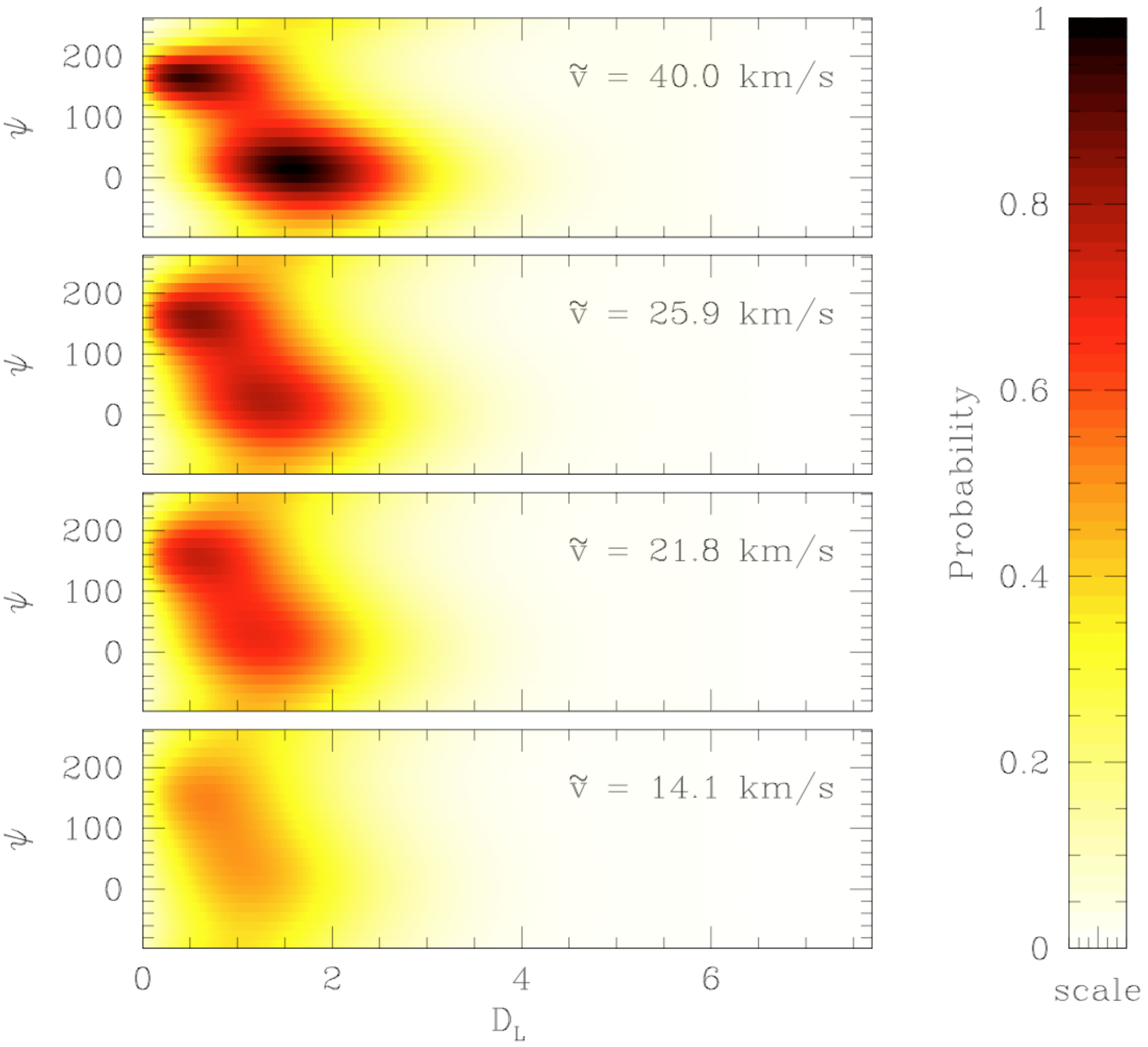}
\caption{The relative $\vpbold$ probability from our Galactic model is plotted
as a function of the lens distance, $D_L$ and angle for four different
values of the projected velocity amplitude, $\vp$, that are representative
of the values that are consistent with the MOA-2007-BLG-192 light curve.
$\psi$ is the angle between $\vpbold$ and the direction of Galactic rotation.
\label{fig-vtil}}
\end{figure}

These values are so similar because a star with the $\vp$ value
of the 2nd best fits is likely to be at a somewhat greater distance
than a star with the $\vp$ value corresponding to the best fits.
This can be understood with the help of Figure~\ref{fig-vtil}, which
shows the results of our Galactic model calculations for the 
probability of $\vpbold$ values as a function of the lens distance
(using heliocentric coordinates).
In this figure, $\vpbold$ is represented by four panels corresponding
to representative projected velocity amplitudes, $\vp$, and the angle
of $\vpbold$ with respect to the direction of Galactic rotation.
The best fits gives $\vp = 21.1\,{\rm km\,s}^{-1}$ and $\psi = 95^\circ$ (from
the direction of Galactic rotation), whereas the 2nd best fits gives
$\vp = 25.0\,{\rm km\,s}^{-1}$ and $\psi \simeq 10^\circ$. So, the 2nd panel from the
bottom of Figure~\ref{fig-vtil} is the one most appropriate to the best 
fit solution, and the third panel from the bottom is the one most
appropriate to the 2nd best solution. The top and bottom panels are
meant to represent the extremes of $\vp$ that are still consistent with
the light curve. 

Note that Figure~\ref{fig-vtil} focuses on the $\vp$ values that are
relevant for the analysis of this event. The 
peak in $\vp$ distribution as determined only by Galactic model 
considerations is at much larger values due to Galactic bulge
lenses. But, these are mostly due to events of relatively short
duration, which have a very low planet detection efficiency. All six of the
planets detected by microlensing of main sequence source 
stars were detected in events
with durations more than 2.5 times longer than the measured
median event timescale of 
$t_E = 16\,$days \citep{macho-bulge-diff,moa-tau,ogle2_blg_tau}.
(This is the median after correction for the event detection efficiency.) 
So, it appears that selection effects imply that events with detected
planets are likely to be much longer than average, which in turn
implies that their $\vp$ values will usually be much smaller than
average. 

We can see that this is
consistent with the likelihood functions of Figure~\ref{fig-masslike} since 
the $\vp = 25\,{\rm km\,s}^{-1}$ and $\psi \simeq 10^\circ$ does favor a larger $D_L$
value than $\vp = 21\,{\rm km\,s}^{-1}$ and $\psi = 95^\circ$. The physical reason
for this can be understood quite simply. Since the average bulge source
star is at rest, the average motion of the observer-source line-of-sight
is that of a rigid body rotating with the Sun's Galactic orbit. The flat
rotation curve of the Galactic disk means that the stars interior to the
Sun will be orbiting faster than the observer-source line-of-sight, with
a $\vp$ value that grows with lens distance from the Sun. However,
if the lens is quite close to the Sun, then the fact that the Sun orbits
$23\,{\rm km\,s}^{-1}$ faster than the average nearby star means that there
is an increased probability that $\vpbold$ will point in the anti-rotation
direction.

\begin{figure}
\plotone{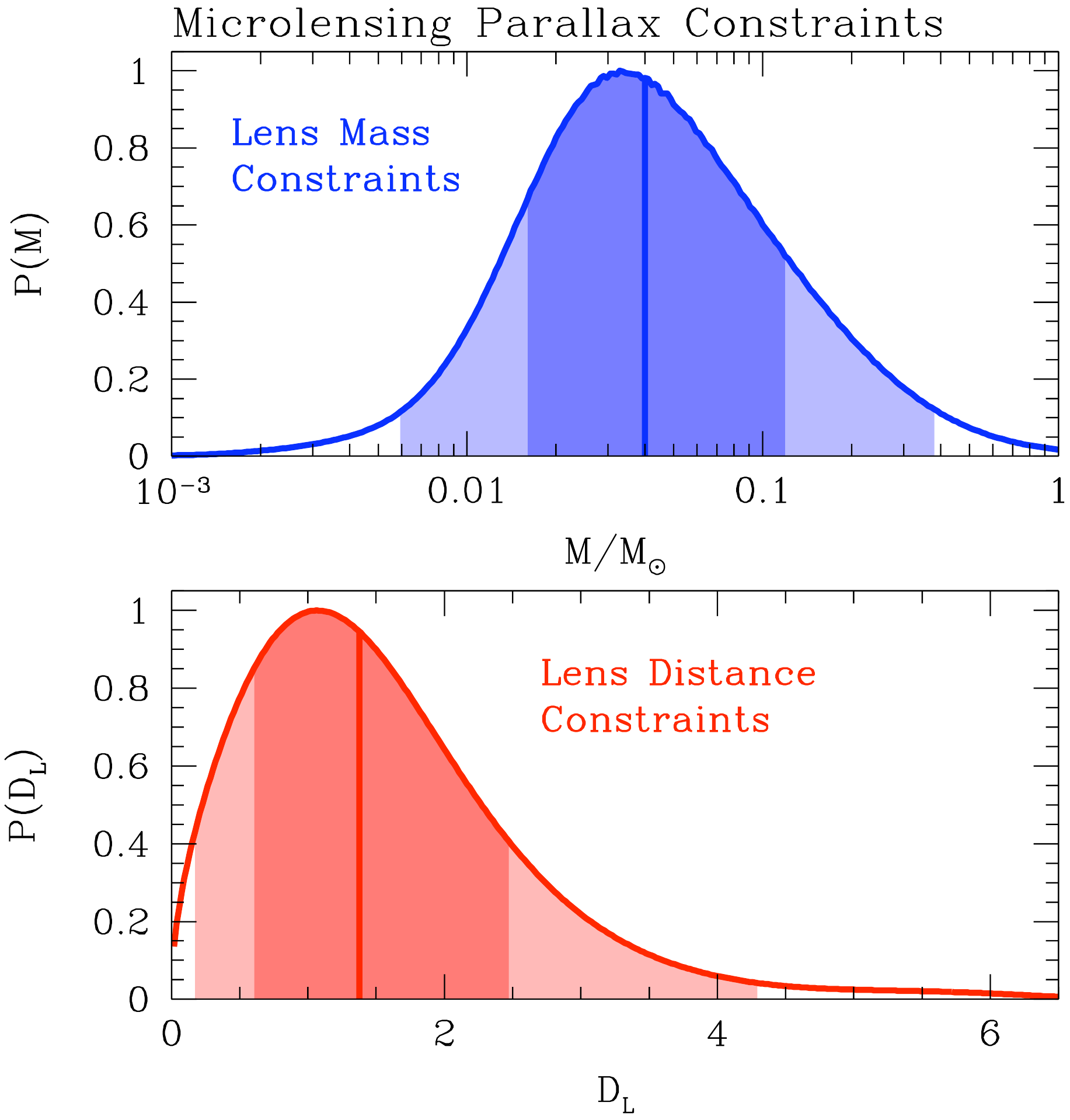}
\caption{Constraints on the lens mass and distance from a Bayesian analysis 
based on the microlensing parallax fits and the Galactic model
described in the text. The lens star mass is 
$M = 0.040{+0.081\atop -0.024}\msun$, with a 2-$\sigma$ range of
0.006--0.39$\,\msun$, and its distance is 
$D_L = 1.4{+1.1\atop -0.8}\,$kpc, with a 2-$\sigma$ range of
0.2--4.3$\,$kpc.
\label{fig-masslike_grid}}
\end{figure}

Of course, Figure~\ref{fig-piEgrid} indicates that there is a range of 
microlensing parallax parameters that are consistent with the 
MOA-2007-BLG-192 light curve, so we can't really base our conclusions
on just the ``best" and ``2nd best" fits. Instead, we can average over the
entire range of fits displayed in Figure~\ref{fig-piEgrid}, with each model
weighted by $e^{-\Delta\chi^2/2}$ compared to the best fit. This gives
the probablity distributions for the lens mass and distance given in
Figure~\ref{fig-masslike_grid}. The results are quite similar to the results
from the individual best and 2nd best fits with 
$M = 0.040{+0.081\atop -0.024}\msun$ and 
$D_L = 1.4{+1.1\atop -0.8}\,$kpc. Thus, at 1-$\sigma$ confidence, the lens
must be a brown dwarf or a late M-dwarf, and at 2-$\sigma$, mid-M dwarfs
would be allowed (although they will not survive the lens brightness 
constraint below).

Although this microlensing parallax analysis depends only upon kinematics
and does not require an input mass function, it is also equivalent to an
analysis including a mass function of the power-law form 
$\Phi \propto M^{-\alpha}$, with $\alpha = 1.5$ \citep{bennett-parbh}. 
Such a mass function
implies that the total lens mass per logarithmic interval decreases with
a power law index of $\alpha -1 = 0.5$, and this is what is needed to
give an equal lensing probability per logarithmic mass interval because
the lensing probability is proportional to the Einstein Radius, 
$R_E \propto M^{0.5}$. If the lens mass function were substantially different
from $\Phi \propto M^{-1.5}$, then the application of a mass prior could 
substantially change the results of the analysis, as is the case for black 
hole lensing \citep{multi-par}. But, this is unlikely to be the case in the
regime of low-mass stars and brown dwarfs. The mass function power-law
index for Galactic disk stars of mass, $M < 0.5\msun$, has been estimated 
to be $0.7 < \alpha < 1.85$ by \citet{kroupa_tout_gil} and 
$1.1 \leq \alpha \leq 1.3$ by \citet{reid_mdwarf_kin}. There are some indications
that $\alpha$ might decrease further in the sub-stellar mass regime
\citep{martin2000,chab2003}, but this change is not dramatic. The 
power-law index could drop to the range $\alpha \sim 0.3$--0.5 for
brown dwarfs. Thus, if we did add a mass function prior to this
analysis, it would not change the relative brown dwarf to low-mass star
probability ratio by much more than a factor of two.

\subsection{VLT NACO Observations}
\label{sec-naco}
\begin{figure}
\plotone{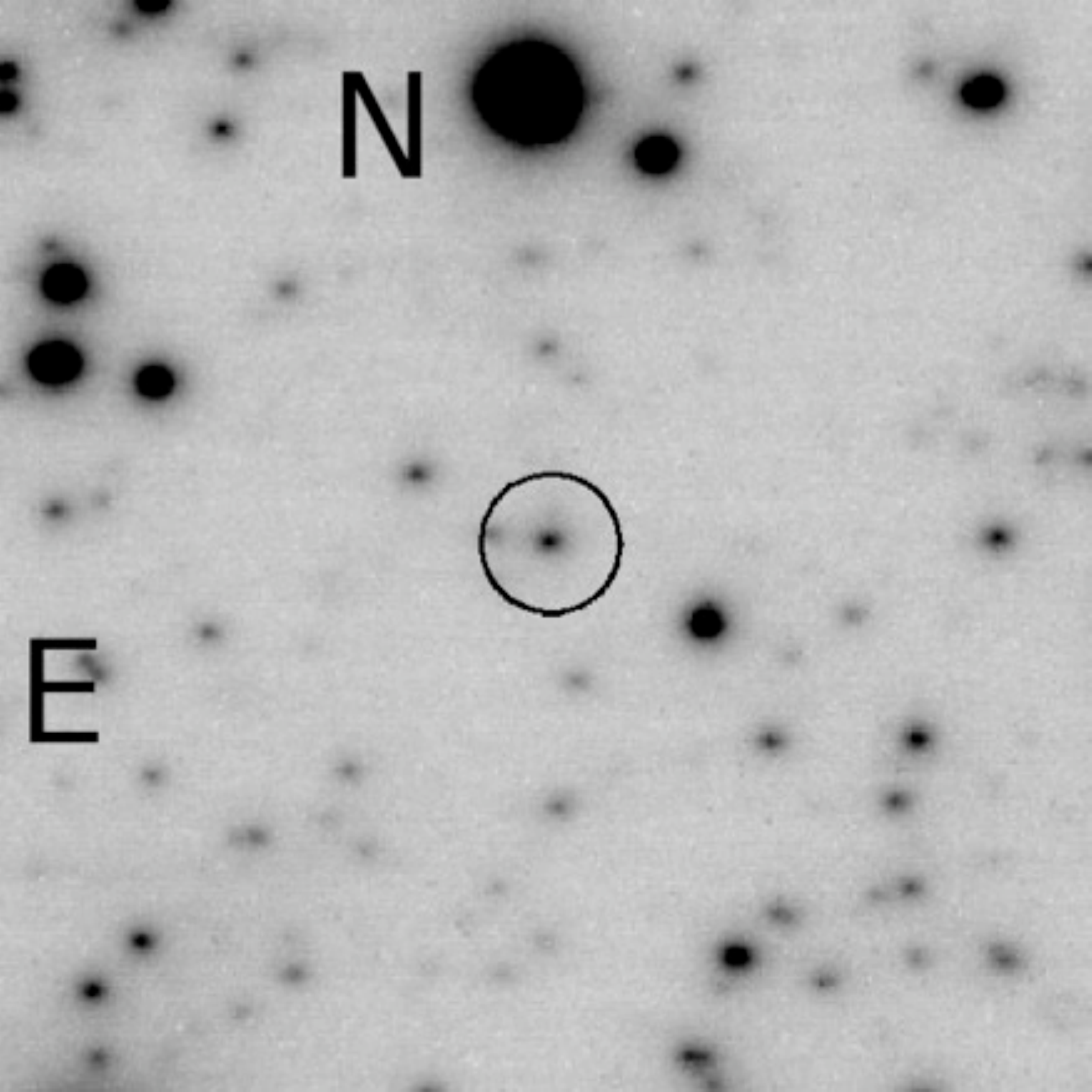}
\caption{A $8.2^{\prime\prime}\times 8.2^{\prime\prime}$ NACO J-band image centered
on the position of the MOA-2007-BLG-192 microlensing event. The source star
is at the center of the black circle near the middle of the image.
\label{fig-NACOimg}}
\end{figure}

A final constraint on the lens star comes from high angular resolution
images taken with the Very
Large Telescope (VLT) using the NACO instrument
on 7 Sep 2007, when the microlensing magnification was $0.23\pm 0.02$ 
magnitudes. The NACO J-band image of the MOA-2007-BLG-192 source has
a point-spread function with a FWHM of $0.15^{\prime\prime}$. It
is shown in Figure~\ref{fig-NACOimg}. A comparison of the stellar positions
in this image with the positions from MOA and OGLE difference images
uniquely identifies the star at the center of the circle in this image as
the source star.

Based upon our measured $I_s = 21.44\pm 0.08$ brightness of the
source star and our estimated extinction ($A_I = 1.61\pm 0.10$), 
we predict that the lensed source
should have $J = 19.69\pm 0.30$ at the time of the NACO images. This 
compares to our measurement of $J = 19.01\pm 0.20$ from our 
preliminary analysis of the NACO data. This uncertainty is dominated by
the photometry zero-point uncertainties due to the small number of
2MASS stars that are seen in the NACO images
The difference between these
magnitudes is $J_b = 19.84 \pm 0.59$. (Here the magnitude error bar
is meant to be interpreted as a linear flux error bar of 
$0.59\times 0.4\ln(10) = 0.54$ times the estimated flux.) Thus,
this measurement is consistent at 2-$\sigma$ with all the J-band flux
coming from the source star, with negligible flux from the lens star.
The possibility of flux from the lens star would appear to be slightly preferred,
but this depends on our uncertain extrapolation from the measured $I$-band
source flux to the $J$-band observations at high angular resolution. It is possible
that we have underestimated the uncertainties in this extrapolation.

If we assume that the excess flux does come from the lens star, then we
can employ a mass-luminosity relation along with equation~(\ref{eq-m_piE})
to determine the lens star mass. There are several mass-luminosity relations
for low-mass stars in the literature to chose from \citep{henry_mc,kroupa_tout,delfosse}.
\citet{henry_mc} offer convenient analytic formulae, and \citet{delfosse} have
fit to newer, more precise data on low-mass stars. But, these formulae both
have somewhat peculiar features at masses where there are few observational
data. The \citet{henry_mc} formulae have discontinuous slopes at the masses where
they chose to change functional forms, and the \citet{delfosse} formulae don't all
extend to the bottom of the main sequence. Therefore, we use the \citet{delfosse} 
mass-luminosity relations for masses in the range 0.12--0.54$\,\msun$, and the 
\citet{henry_mc} relation for $M < 0.10\msun$. For masses in the 0.10--0.12$\,\msun$
range, we linearly interpolate between the two. The $J$-band mass-luminosity relation
is nearly indistinguishable from the \citet{kroupa_tout} for masses below $0.25\msun$

If we apply the constraint $J_L = 19.84\pm 0.59$ on the lens star, the fit is driven to
a smaller source radius crossing time than the best fit. This constrained
lens brightness fit has $t_\ast = 0.035\,$days, 
and a lens star mass of $M = 0.092\msun$. The implied distance for the lens star
is $D_L =  0.55\,$kpc, and the fit $\chi^2 = 1116.63$, which is only 
$\Delta\chi^2 = 1.17$ worse than the best fit cusp approach solution.
The caustic crossing solutions do not allow a bright lens star
because they require $t_\ast > 0.08\,$days, which keeps the lens mass below
the Hydrogen-burning threshold of $0.08\msun$.
As indicated in Figure~\ref{fig-Ts_eps_grid}, the low-$t_\ast$ 
cusp approach solutions
prefer a small mass ratio, $q\sim 4\times 10^{-5}$. This best fit, bright lens
solution has $q = 4.5\times 10^{-5}$, which implies a planetary mass of only
$m = 1.4\mearth$. 
Thus, if the lens star is at the bottom of the main sequence,
the planet's mass is likely to be even lower than implied by the best light curve
models, which would imply a brown dwarf planetary host.

Follow-up VLT/NACO
images in 2008 should generate much more precise limits on the lens
and source star magnitudes and colors.
If the lens star is not a brown dwarf, follow-up Hubble Space
Telescope (HST) images will detect the lens-source relative proper motion, which will
significantly reduce the uncertainty in the lens parameters by measuring
$\phi_E$ and the lens-source relative proper motion 
$\mu_{\rm rel} = \theta_\ast/t_\ast$ \citep{ml_plan_char}. For large lens
masses, eq.~(\ref{eq-m_piE}) implies that $D_L \propto 1/M$, so even a 
white dwarf primary as old as the Galactic disk would be bright enough 
\citep{hansen_ngc6397} to detect in HST images. If follow-up HST images
cannot detect the lens primary in the $V$ and $I$-bands, we could conclude
that the planetary
host star must be a brown dwarf as the best fit light curve model indicates.

\subsection{Combined Parallax, Finite Source, and Lens Brightness Constraints}
\label{sec-par_fin}

We have now explored the degeneracies in the light curve model that allow the
16 local $\chi^2$ minima presented in Table~\ref{tab-fitpar}, and we have also
considered the constraints that can be put on the planetary host
star mass and distance from microlensing parallax, finite source effects, and the
possible detection of the lens in adaptive optics images from the VLT/NACO
instrument.
These constraints can now be combined in a MCMC analysis. We have run
16 independent MCMC runs centered on each of the 16 local minima listed
in Table~\ref{tab-fitpar}, and each of these runs has been subject to the
constraint on the $J$-band brightness of the lens, $J_L \geq 19.84\pm 0.59$.
Additional constraints were added to some of the runs to prevent them from
passing from the region of one local minimum to another, and each chain
had approximately 40,000 steps. Because of poor sampling of the $q > 10^{-3}$
regions in the cusp approach solutions, we have also included 8 additional
MCMC runs that have not reached equilibrium that are intended to sample
this region of parameter space.
The $q$ vs. $d$ distribution from these combined MCMC runs is plotted in
Figure~\ref{fig-mcmc}.

In order to combine the results of the MCMC runs in the regions of
these different local minima, we must weight each MCMC chain by the
$e^{-\Delta\chi^2/2}$ factor for the $\chi^2$ value of the relevant local 
minimum. For the large $q$ non-equilibrium runs, we try a slightly different
approach (suggested by A. Gould). We have done a high temperature 
MCMC run in the vicinity
of solution A from Table~\ref{tab-fitpar} with three times the normal
temperature (so the Boltzman probability factor, $e^{-\Delta\chi^2/2}$
is replaced by $e^{-\Delta\chi^2/6}$). Then, we calculate the ratio
of the sums of $e^{-\Delta\chi^2/2}$ for all the points in the high-$q$
region to the region of parameter space that is well sampled by
the normal temperature MCMC runs. This ratio is then used as 
a relative weighting to apply to these out of equilibrium runs.
We find a ratio of $8\times 10^{-4}$ for the solutions with $q > 10^{-3}$
and $t_0 < 4245.43$. However, this procedure is ambiguous because
the high temperature MCMC runs cover parts of parameter space not
covered by the low temperature runs, so this ratio depends on the
precise boundaries used for the high-$q$ region. Different choices
can easily change the weighting for the high-$q$ region by a factor
of two. Fortunately, this uncertainty has no influence on our final
estimates of the lens properties.

Because each model provides microlensing parallax
parameters that allow us to determine $D_L$ and $\vpbold$, we can
also apply a prior probability for each model based upon its 
likelihood in our assumed Galactic model. We have computed
these probabilities on a grid in $\vp$, $\psi$, and $D_L$ with $\vp$ ranging
from $10\,{\rm km\,s}^{-1}$ to $61.7\,{\rm km\,s}^{-1}$ in logarithmic intervals of $2^{1/8}$,
$\psi$ ranging from 0--$360^\circ$ in $15^\circ$ intervals, 
and $D_L$ ranging from 0 to $7.7\,$kpc in
$38\,$pc intervals. Figure~\ref{fig-vtil} shows four $\vp =\,$constant 
slices of this probability distribution.
This range in (heliocentric) $\vp$ is sufficient to cover
all of the $\vp$ values that occur in the MCMC runs. The probabilities
to be used in the MCMC parameter estimates are determined by
interpolation from this grid.

\begin{figure}
\plotone{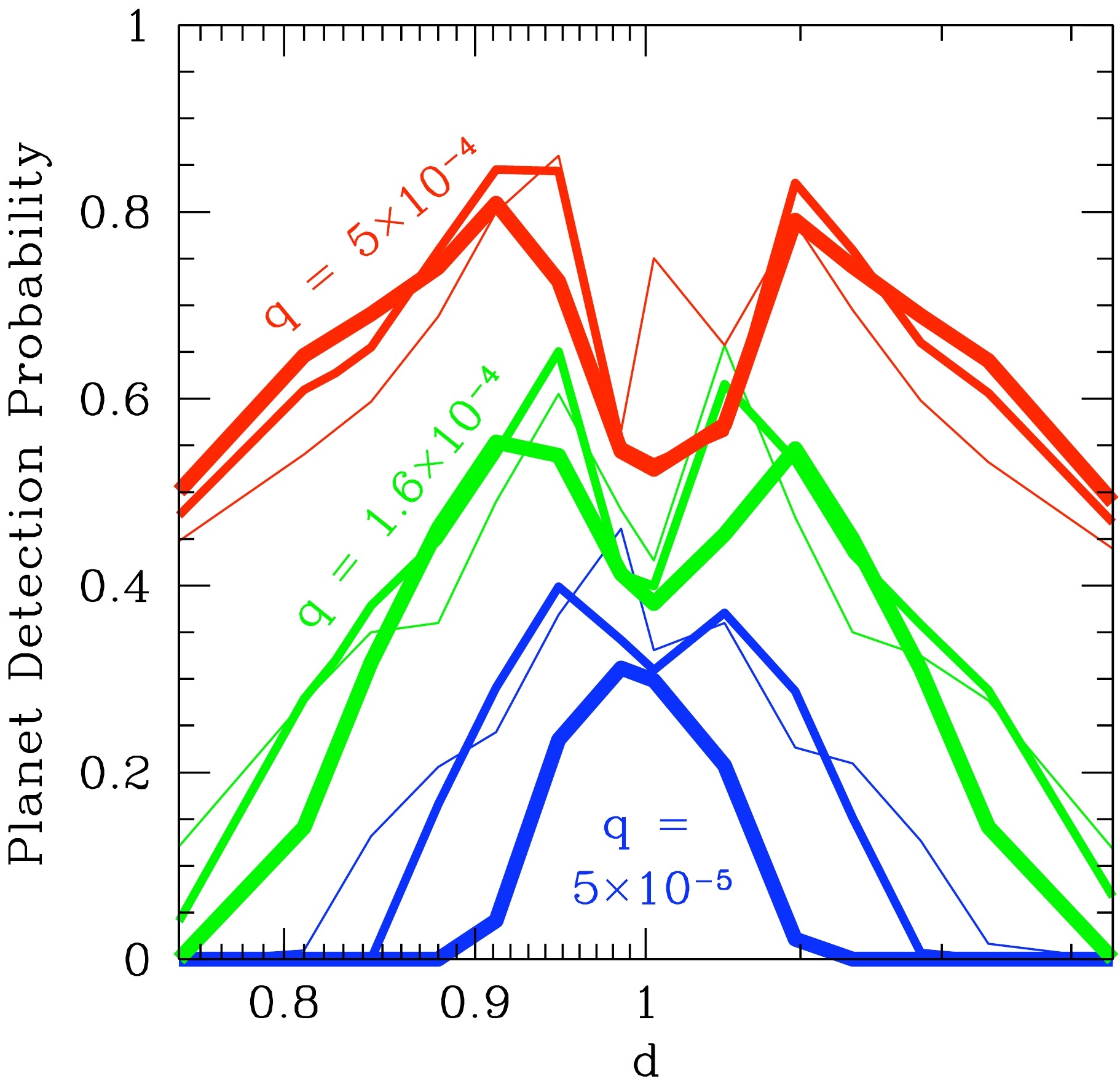}
\caption{
The planet detection probability for MOA-2007-BLG-192 is plotted 
as a function of the separation, $d$, for mass ratios of 
$q = 5\times 10^{-5}$, $1.6\times 10^{-4}$, and $5\times 10^{-4}$
is blue, green, and red, respectively. The thin, medium, and thick
curves represent source star crossing times of $t_\ast = 0$, 0.06428,
and $0.11353\,$days, respectively.
\label{fig-det_prob}}
\end{figure}

An important issue is whether to impose any other prior to the models
when summing over the MCMC results to estimate parameter probability
distributions. We do not believe that the stellar or planetary mass functions
are well known enough to impose any prior distribution on them, nor do
most of the other model parameters warrant a prior probability distribution.
The one exception is the lens separation, $d$. Because of this event's
high magnification and the fact that the solutions we consider all have
$0.8 \simlt d \simlt 1.25$, we expect that the true
distribution in $d$ is relatively flat across the region where microlensing
is sensitive. This expectation is borne out by the explicit calculation
presented in Figure~\ref{fig-det_prob}, which presents the planet detection
probability calculated with the method of \citet{mps-98blg35} using
a detection threshold of $\Delta\chi^2 \geq 320$. This figure shows some $d$
dependence, but we are interested in the $d$ ranges of $0.9 \simlt d \simlt 1.1$
for $q =  5\times 10^{-5}$, $0.85 \simlt d \simlt 1.18$ for 
$q = 1.6\times 10^{-4}$, and $0.77 \simlt d \simlt 1.3$ for
$q =  5\times 10^{-4}$. Also, the largest value of $t_\ast$ shown in
Figure~\ref{fig-det_prob} is only relevant for the caustic crossing 
solutions at $0.98  \simlt d \simlt 1.01$ and not the cusp approach
solutions with $d$ further from 1. Thus, it is a reasonable approximation
to neglect the $d$ dependence of the detection probability.

The caustic crossing solutions require that the
lens be located very close to the Einstein ring--within the range
$0.98 \simlt d \simlt 1.01$. This contrasts to the much larger ranges
in $d$ that allow the cusp approach solutions: $0.75 \simlt  d \simlt 0.95$
and $1.05 \simlt  d \simlt 1.3$. Thus, the {\it a priori} probability of 
the caustic crossing solutions would appear to be 
substantially lower than the probability of the cusp crossing solutions.

\begin{figure}
\plotone{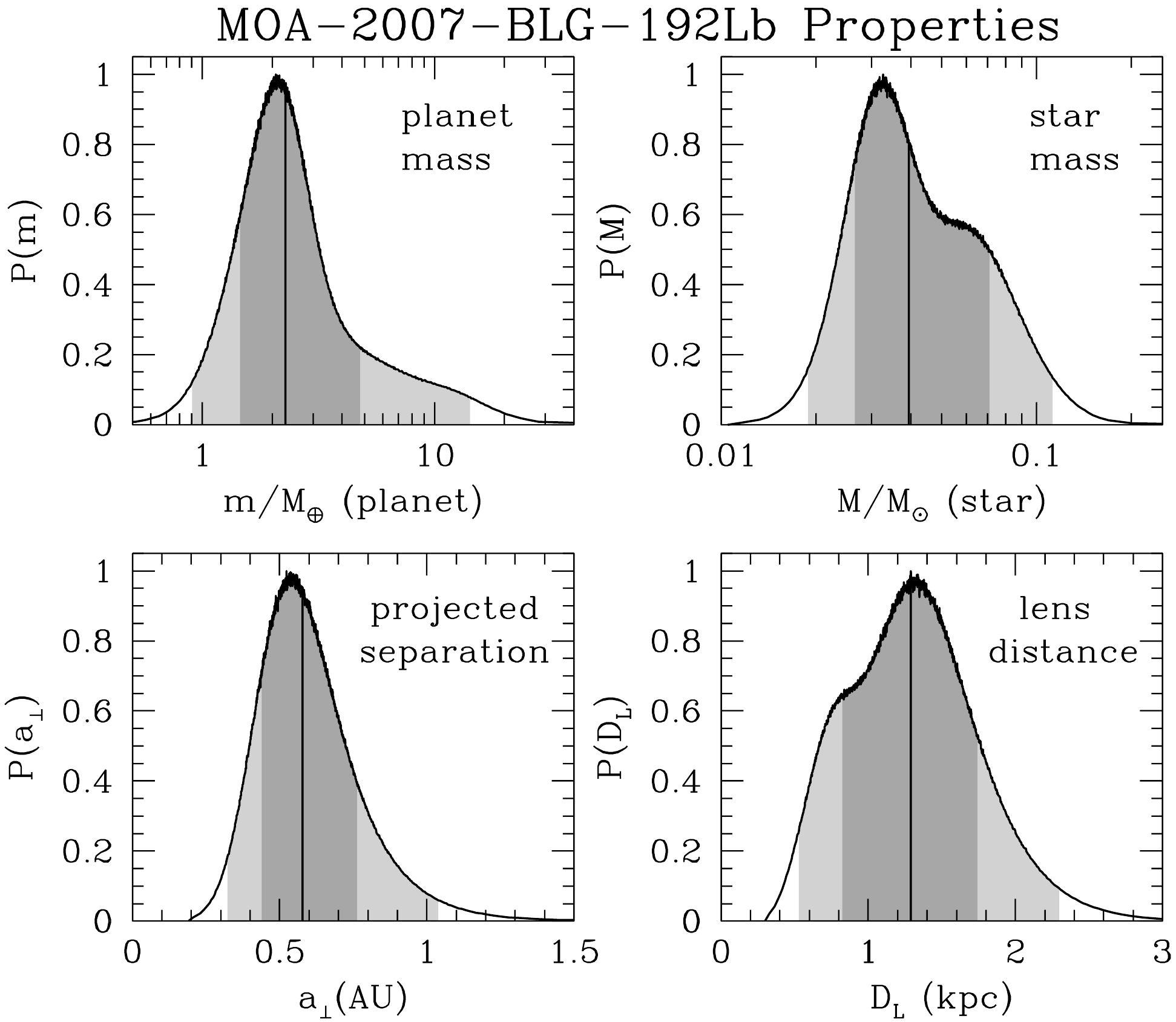}
\caption{The relative probability distributions for the planet and star
masses, the planet-star projected separation, and the distance to the
lens system are plotted for the MOA-2007-BLG-192 event
without employing any prior
probability distribution on the star-planet separation, $d$. Each plot
is constructed from the combined 16 MCMC runs corresponding to each of the 
local minima listed in Table~\ref{tab-fitpar} plus the 8 additional large $q$ runs
as discussed in the text. The dark and light grey shaded
regions indicate the 1-$\sigma$ and 2-$\sigma$ confidence intervals, respectively,
and the vertical lines indicate the median of each distribution. These values are
also reported in Table~\ref{tab-puncert}.
\label{fig-lens_prop}}
\end{figure}

\begin{figure}
\plotone{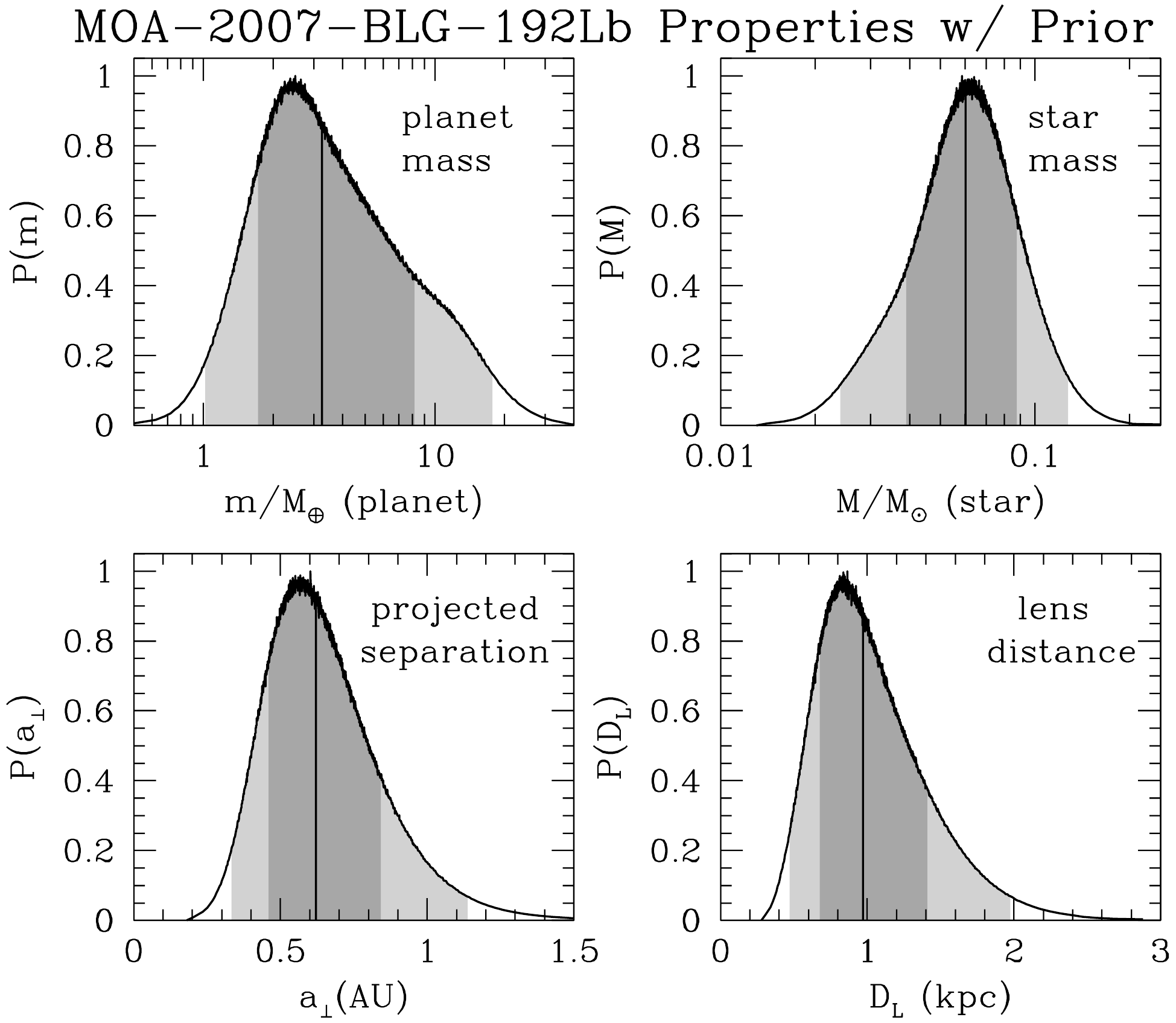}
\caption{The relative probability distributions for the planet and star
masses, the planet-star projected separation, and the distance to the
lens system are plotted for the MOA-2007-BLG-192 event
with a prior on the $d$ distribution favoring the cusp approach solutions by a factor
of 10. Each plot
is constructed from the combined 16 MCMC runs corresponding to each of the 
local minima listed in Table~\ref{tab-fitpar} plus the 8 additional large $q$ runs
as discussed in the text. The dark and light grey shaded
regions indicate the 1-$\sigma$ and 2-$\sigma$ confidence intervals, respectively,
and the vertical lines indicate the median of each distribution. These values are
also reported in Table~\ref{tab-ppuncert}.
\label{fig-lens_prop_pri}}
\end{figure}

However, the question of whether to employ a prior on $d$ is somewhat
subtle. One might regard the narrow range of $d$ that allows caustic crossing
solutions as simply a feature of the very high sensitivity of the light curve
shape to changes in the parameters for $d\approx 1$. This would account
for the very narrow allowed range of $d$ values for the caustic crossing 
solution. On the other hand, $d\approx 1$ is a very special region that 
corresponds to many unusual features in planetary light curves, and it is
much more likely for a planet to be located in the allowed ranges for the 
cusp approach solutions than for the caustic crossing solutions. So, we
favor applying a prior that favors the cusp approach solutions by a factor 
of 10:1 over the caustic crossing solutions. The probability distributions
resulting from a Bayesian analysis over all 16 of the MCMC runs without
this prior are given in Figure~\ref{fig-lens_prop}, while the probability
distributions with this prior are given in Figure~\ref{fig-lens_prop_pri}.
The median, 1-$\sigma$ and 2-$\sigma$ uncertainties are given in
Table~\ref{tab-puncert} without the prior and Table~\ref{tab-ppuncert}
with the prior.

The main difference in the parameter distributions with and without the
$d$ prior is that the higher weighting of the caustic crossing solutions
without the prior pushes the lens primary mass lower, from 
$M = 0.060{+0.028\atop -0.021} \msun$ to
$M = 0.042{+0.021\atop -0.015} \msun$. The planetary mass
estimate also drops by a similar factor, from
$m = 3.3{+4.9\atop -1.6} \mearth$ to
$m = 2.3{+2.3\atop -1.2} \mearth$. This is a consequence of the 
larger $t_\ast$ value for the caustic crossing solutions, which is the
only important difference between these solutions. So, the qualitative 
conclusions are similar whether or not we use the $d$ distribution
prior.

One might expect that the distributions in $a_\perp$ might be 
bimodal in Figure~\ref{fig-lens_prop_pri}, since the caustic crossing 
solutions at $d \simeq 1$ have a low weight. However, the distributions
in $t_\ast$, $\pi_E$, and $\theta_\ast$ are broad enough to remove the
local minimum that we might expect due to the $d \leftrightarrow 1/d$
ambiguity.

\section{Discussion and Conclusions}
\label{sec-conclude}

MOA-2007-BLG-192 is the first planetary microlensing event to be
discovered without follow-up observations of the light curve, and this
discovery was made possible by the very wide field (2.2 sq. deg.)
of the MOA-II telescope, which allows the entire Galactic bulge
to be imaged hourly. This hourly coverage is sufficient to establish
that the light curve deviation can only come from a planet and not
a stellar binary, it does leave the planetary parameters less well
constrained than would be the case for a high magnification
event that had been discovered and announced significantly before
peak magnification. As a result, there is significant uncertainty in
some of the event parameters, including the source radius crossing
time, $t_\ast$, and the planetary mass ratio, $q$. Some of this uncertainty
is due to two different types of light curve models that can explain the data:
a cusp approach model and a caustic crossing model.

However, this event also has a significant microlensing parallax signal,
which indicates that the planetary host star is likely to have a very low
mass, with sub-stellar masses favored. A sub-stellar planetary host
mass is favored even more when the marginal measurement of $t_\ast$
is included in the analysis. However, a preliminary analysis of AO
imaging from the VLT/NACO instrument indicates that there may be
some excess stellar $J$-band flux at the location of the source star,
although the significance of this excess is less than 2-$\sigma$.

If this $J$-band excess is due to flux from the lens star, then it
can be explained by a cusp approach model that has a 
$\chi^2$ value larger than the best fit by $\Delta\chi^2 = 1.2$.
The implied planet host star mass is $M \approx 0.09\msun$, and the implied
planet mass is quite low, $m \approx 1.4\mearth$ because the best low
$t_\ast$ solutions that correspond to stellar mass host stars also have
low planet mass ratios.

If the host star is a low-mass star instead of a brown dwarf, then we will have
a further opportunity to improve our characterization of the lens system
with follow-up HST observations \citep{ml_plan_char}. These stellar mass
lens models require a lens-source relative proper motion of
$\mu_{\rm rel} \simgt 5\,$mas/yr, and this is enough to allow the
detection of the separation in HST images taken in a few years.
Although the lens and source stars will not be resolved, their separation
will be large enough so that the blended lens+source image will be elongated,
and the very stable PSF of HST will allow this elongation to be measured
precisely when the separation is $\simgt 15\,$mas. Such a measurement,
will pin down the direction of lens-source motion, which will help to 
restrict the remaining uncertainty in the microlensing parallax measurement.
This would dramatically reduce the parameter uncertainties for this event
and leave only one parameter, the mass ratio $q$, that will have a substantial 
uncertainty (although the uncertainty in $q$ will be slightly reduced with
a more precise measurement of $t_\ast = \theta_\ast/\mu_{\rm rel} $).

The only possibility that could seriously modify our main conclusion of a very
low planetary host star mass would be if the apparent microlensing parallax 
signal were due to xallarap (or orbital motion of the source star). However, we
argue in the Appendix that this is unlikely.

Our analysis indicates that the planetary host star, MOA-2007-BLG-192L, is likely to
have a mass in the 0.02--$0.10\,\msun$ range (at 95\% confidence), 
and it is the first such object known
to have a companion with a planetary mass ratio ($q < 0.03$). Two brown
dwarfs have previously been reported to have companions that could be of
planetary mass \citep{chauvin,chaha8}, but their mass ratios are $q \sim 0.2$,
which suggests that they did not form like the planetary systems around more
massive stars. Thus, our discovery of MOA-2007-BLG-192Lb represents the
first discovery of an extrasolar planet with a planetary mass ratio orbiting
an extremely low-mass primary that is likely to be a brown dwarf, and with
a mass of $m = 3.3{+4.9\atop -1.6} \mearth$ (or a 2-$\sigma$ range of
1.0--$18\,\mearth$). Thus, the median estimated mass for
MOA-2007-BLG-192Lb represents the lowest
mass  for a planetary companion yet to be discovered, aside from the 
lowest mass of the pulsar planets
\citep{pulsar_planets}. This discovery suggests that planetary systems
can form around stars of extremely low-mass \citep{payne_BD_planets},
and confirms that microlensing is indeed sensitive to Earth-mass
planets \citep{em_planet}. It also indicates that Earth-mass planets should be
able to form around very low-mass M-dwarfs, which should provide
encouragement for programs that seek to find transiting Earth-mass planets
in the habitable zone of M-dwarfs in order to study their atmospheres.

In fact, it is possible that MOA-2007-BLG-192Lb could have
a habitable surface temperature itself, despite the fact that its
host star or brown dwarf provides extremely feeble radiative heating. 
\citet{stevenson99} has speculated that even a free floating 
Earth-mass planet could have a surface temperature that would allow
liquid water even though the heating from internal
radioactive decays provides a factor of $\sim 10^4$ times less
energy than the Earth receives from the Sun. The key point of 
Stevenson's argument was that such a free floating planet might
retain a molecular Hydrogen atmosphere that could provide very
strong insulation that would allow the surface temperature to
remain above the melting point of water ice. If it was possible to
detect nearby analogs to MOA-2007-BLG-192Lb, it would be 
worthwhile to attempt to study their spectra to see if they do
have $H_2$ atmospheres that might allow warm surface
temperatures.

\begin{figure}
\plotone{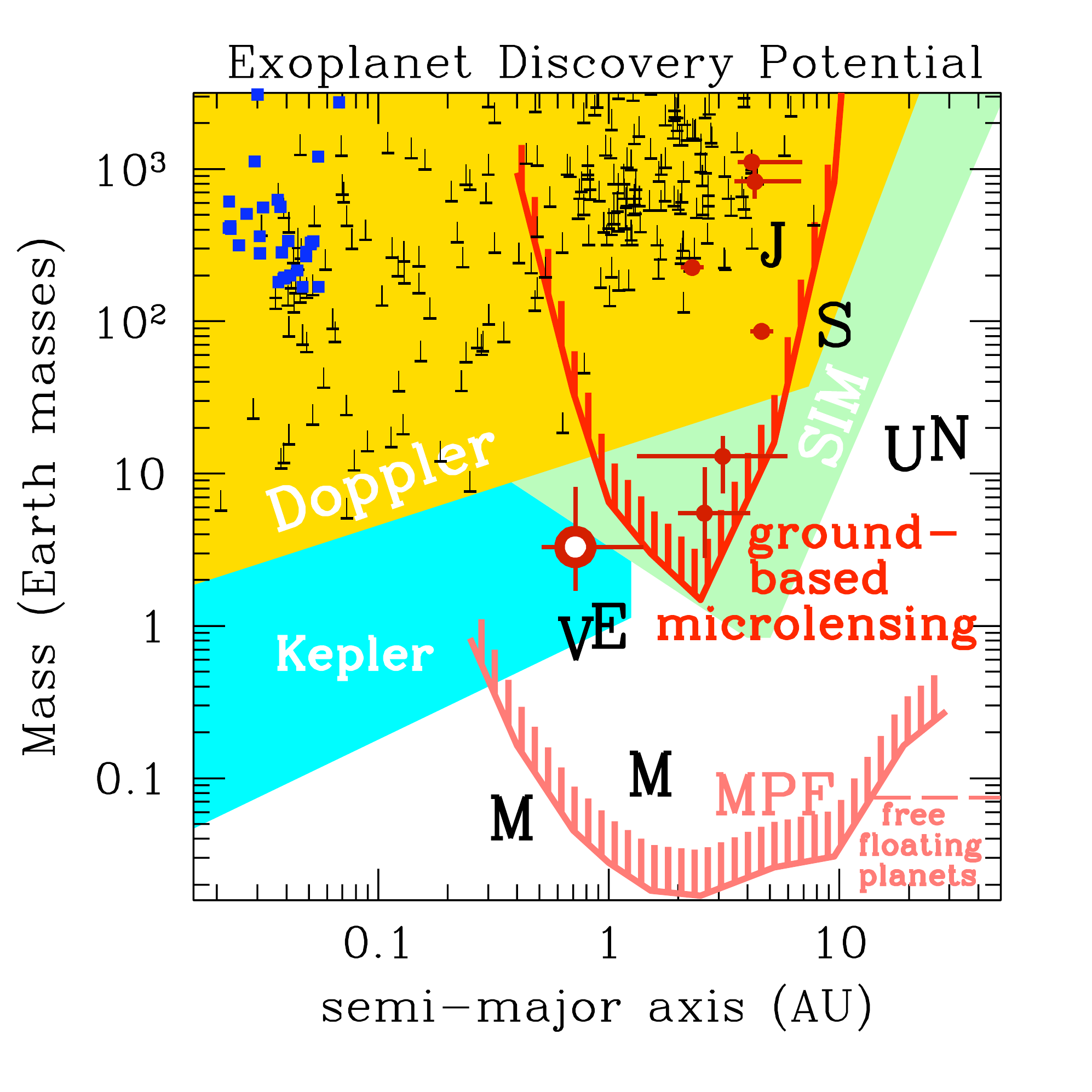}
\caption{The known extrasolar planets are plotted as a
function of mass vs. semi-major axis, along with the predicted 
sensitivity curves for a number of methods. The microlensing planets
are indicated by dark red spots with error bars, and the large red spot
with a white dot in the center is MOA-2007-BLG-192Lb.
The blue dots indicate
the planets first detected via transits, and the black bars with upward
pointing error bars are the radial velocity planet detections. (The upward
error bars indicate the 1-$\sigma$ $\sin i$ uncertainty.) The gold, cyan,
and light green shaded regions indicated the expected sensitivity of the 
radial velocity programs and the Kepler and SIM space missions. The
dark and light red curves indicate the predicted lower sensitivity limits
for a ground based and space-based \citep{gest-sim} microlensing
planet search program, respectively. The Solar System's planets are
indicated with black letters.
\label{fig-mass_a}}
\end{figure}

Figure~\ref{fig-mass_a} compares this new discovery (indicated by the
red circle surrounding a white spot) to previous discoveries by microlensing
and other methods. We should note that this discovery appears outside the
predicted sensitivity range for ``ground-based microlensing\rlap." This is
largely because the ground-based calculations were done for planets with
fixed mass ratios and separations, and they were added to this plot 
using the assumption that the ``typical" planetary host mass was
$0.3\msun$. On a mass ratio plot, the MOA-2007-BLG-192Lb would appear
much higher and closer to the ``ground-based microlensing" curve.

\begin{figure}
\plotone{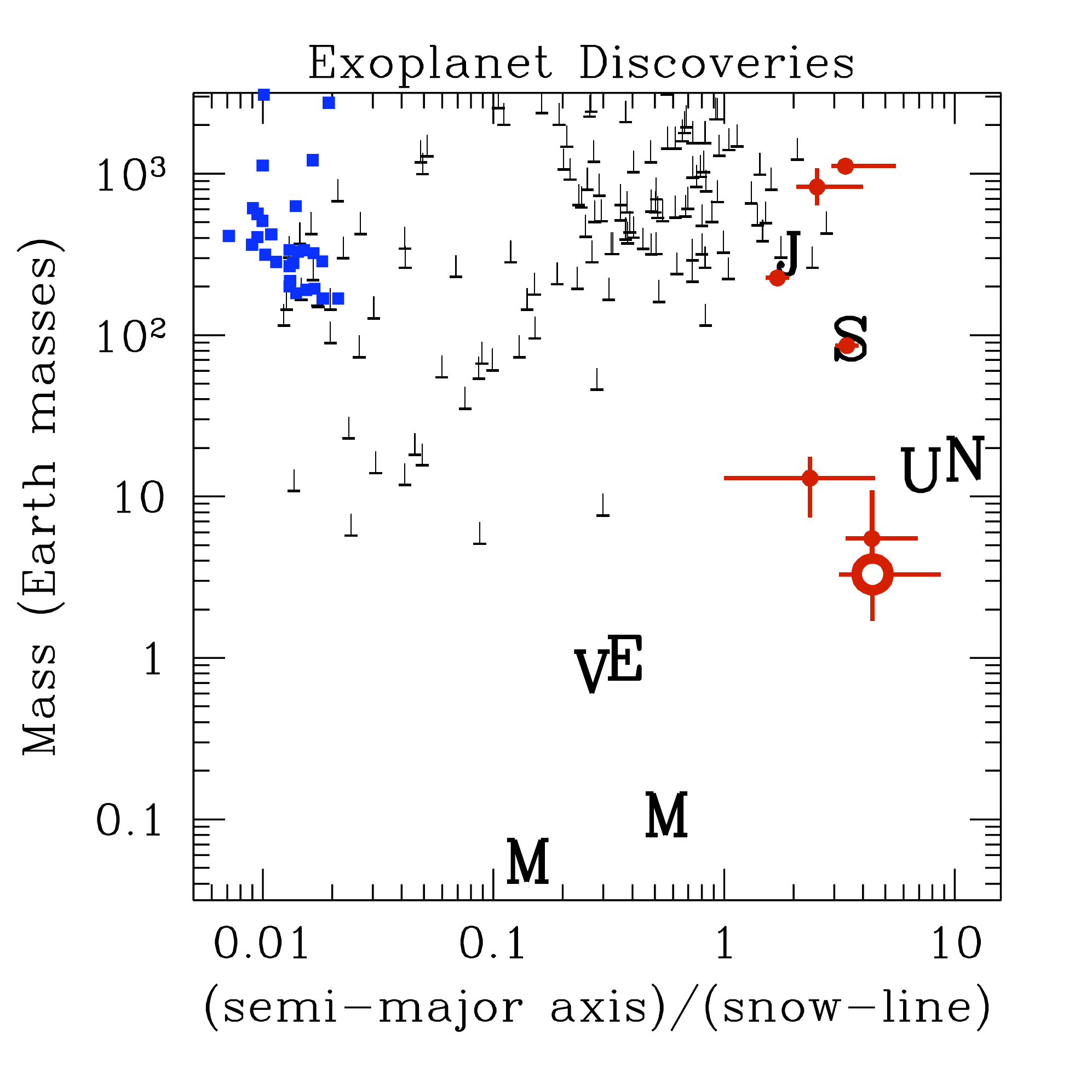}
\caption{The known extrasolar planets are plotted as a function of their
mass and semi-major axis divided by the snow line, which is taken to 
be at $a_{\rm snow} = 2.7\,{\rm AU}\,M/\msun$. As in Figure~\ref{fig-mass_a},
microlensing planets
are indicated by dark red spots with error bars, and the large red spot
with a white dot in the center is MOA-2007-BLG-192Lb. Blue dots indicate
the planets first detected via transits, and the black bars with upward
pointing error bars are the radial velocity planet detections.
\label{fig-mass_snow}}
\end{figure}

In this plot, our new planet looks as if it might be a more massive version
of Venus, but this is a bit misleading since such a low mass primary 
provides very much less heating than the Sun. If we want to consider the
planet locations in the context of planet formation, then it is best to
consider not the planetary semi-major axis, but the planetary semi-major
axis divided by the ``snow-line". In the context of planet formation theory
\citep{ida_lin,lecar_snowline,kennedy-searth,kennedy_snowline}, 
the ``snow-line" is the region in 
the proto-planetary disk where it is just cold enough for water-ice to form.
This is expected to increase the density of solids in the disk by a factor of
$\sim 5$, and it is where the most massive planets are expected to form,
according to the core accretion theory. Since planets are expected to form
early in a star's history, it is not the star's main sequence brightness that 
determines the location of the snow line. Instead it is the star's brightness
at an age of $\sim 1\,$million years, when the stellar luminosity is thought to
scale as $\sim M^2$
\citep{burrows93,burrows97} (G. Kennedy, C. Lada, private communications).
Thus, we can estimate the distance of the ``snow-line"
to be $a_{\rm snow} = 2.7 M/\msun$, and with this definition we plot the
known exoplanets as a function of mass and (semi-major axis)/(snow-line)
in Figure~\ref{fig-mass_snow}. Now with respect to the snow-line, 
MOA-2007-BLG-192Lb appears to be a much lower mass version of
Uranus instead of a massive Venus. This figure also shows that, to
date, only microlensing has been able to probe the region beyond the
snow line, for planets of less than a Jupiter mass. Our discovery adds
strength to the claim that low-mass planets are substantially more
common at these separations around stars of less than a Solar mass
\citep{ogle169}.

\acknowledgments
We wish to thank Subo Dong for calculations that pointed out an important error in
the original version of this manuscript. We also thank
the referee, Andy Gould, for a very comprehensive review
that helped to improve the paper in a number of key areas.
We thank Doug Lin, Grant Kennedy, Scott Kenyon and Adam
Burrows for advice on the scaling of the snow-line with stellar mass.
MOA is supported by the Marsden Fund of New Zealand, a Grant-in-Aid for 
Scientific Research on Priority Area (19015005) by the Ministry of
Education, Culture, Sports Science and Technology (MEXT) of Japan, and
three Grants-in-Aid for  for ScientiÞc Research 
(17340074, 18253002 and 18749004) by the
Japan Society for the Promotion of Science (JSPS).
D.P.B.\ was supported by grants 
AST-0708890 from the NSF and NNX07AL71G from NASA. 
The OGLE project was partially supported by the MNiSW grant
N20303032/4275.
S.H. and Y.P. gratefully acknowledge support by Science Faculty 
Scholarships at the University of Auckland.
D.K. and S.B. would like to express their gratitude to ESO, especially C. Dumas,
C. Lidman and S. Mengel  for the succesful execution of the NACO observations of DDT program 279.C-5044(A).

\appendix
\section{Parallax vs. Xallarap}
\label{sec-xal}

It is always the case that the orbital motion of the source star can reproduce the
same light curve as the orbital parallax effect \citep{gould-par1,macho-par1} 
because it is possible for the source star to have a binary orbit that mimics
that of the Earth \citep{smith_par_acc,multi-par}. This is often referred to as 
the ``xallarap" effect, because it is the reverse of parallax.
In high magnification events, like MOA-2007-BLG-192, it
is possible to definitively distinguish between xallarap and parallax by
measuring the terrestrial parallax effect \citep{holz_wald} caused by the
different positions of the telescopes on the surface of the Earth. Indeed, 
the black and blue curves in the lower panel of Figure~\ref{fig-lc} 
show the MOA-2007-BLG-192 light curves as seen from the MOA-II
telescope at Mt. John, New Zealand, and the OGLE telescope at
Las Campanas, Chile. The difference is clearly enough to be measured
with very dense sampling of the light curve that might have been achieved
if the event had been detected and announced earlier. But with the relatively 
sparse sampling of the observed light curve, we make a marginal detection
of the terrestrial parallax effect. The best light curve fit including this effect
has a $\chi^2$ improvement of $\Delta\chi^2 = 0.5$ compared to the best 
fit without this effect. So, terrestrial parallax is detected at 0.7-$\sigma$
significance, which is obviously not enough to exclude the possibility that 
the apparent parallax signal is due to xallarap .

It is also possible to definitively distinguish between parallax and xallarap
by detecting the lens star \citep{bennett-moa53,ml_plan_char,dong_ogle71}
because these models make very different predictions for properties of
the lens star. Parallax generally predicts a lens star that is nearby, while
a xallarap model requires that the lens be more distant  and substantially
more massive (assuming that $\theta_E$ has already been fixed through
the measurement of $t_\ast$ or the lens-source relative proper motion,
$\mu_{\rm rel}$).

With our present data, however, we are not able to definitively exclude
the alternative xallarap explanation of the apparent orbital parallax features
in the light curve. However, we are able to show that the parallax model fits
the data at least as well as the best xallarap models and that the 
xallarap models that are consistent with the data are limited to a very
small region of the possible parameter space. This implies that a 
xallarap model is very unlikely, although not completely excluded.

\subsection{Xallarap Model Fitting}
\label{sec-xal_fit}

Xallarap is most conveniently described with parameters that are
very similar to the parameters that we use for the microlensing 
parallax model. We assume a circular orbit for simplicity, because
the non-circular orbital parameters are unlikely to improve the
fits significantly. The orbital motion of the source affects the 
apparent lens-source relative motion in the same way as in 
a microlensing parallax model, so we can define an analog to the
microlensing parallax vector, $\piEbold$. To avoid confusion, we
will denote the magnitude of the xallarap vector by $\xi_E$, while
the angle by the same $\phi_E$ used for the parallax models.
Unlike the parallax case, we do not know the orientation of the 
source star orbit, so we must include additional parameters to
describe the orientation. For a parallax model the orientation of the
orbit is fixed by the location of the source star in the sky, so we can
describe the orientation of the source star orbit by the position of
the Sun in the pseudo-ecliptic coordinates based on the orbit of
the source star. The pseudo-ecliptic longitude and latitude of the
Sun as seen from the source are denoted by $\lambda_s$ and
$\beta_s$, respectively. The phase of the source star orbit is given by
$\lambda_s$, and $\beta_s$ gives the angle between the 
line-of-sight and the source orbital plane.

We have done an extensive series of xallarap model fits in order
to determine the parameter range of potentially viable models. Some
of these results are shown in
Figure~\ref{fig-xal_grid}, which
indicates the distribution of $\chi^2$ differences
between the best xallarap models with orbital periods of 1-year
and the best parallax models for both the cusp approach and
caustic crossing models. These plots only show 
$180^\circ < \lambda_s \leq 360^\circ$ because there is an
exact symmetry relating the $ \lambda_s \leq 180^\circ$ models
to the $ \lambda_s \leq 360^\circ$ models \citep{dong_ogle71}.
The best models on this grid have $\chi^2$ values
that are slightly worse than the best microlensing parallax models,
with a difference of $\Delta\chi^2 = 0.35$ for the cusp approach models
and $\Delta\chi^2 = 0.25$ for the best caustic crossing models. If we 
don't constrain $\lambda_s$ and $\beta_s$, to these grid points,
we find that the best xallarap solutions are now better by
$\Delta\chi^2 = 0.09$ and $\Delta\chi^2 = 0.20$. If both were
acceptable models, we'd expect the xallarap models to be
better by $\Delta\chi^2 \sim 2$, so parallax is slightly favored
over xallarap based on the light curves alone.

Figure~\ref{fig-xal_grid} indicates that the majority of the $P = 1\,$yr 
parameter space for both the cusp approach and caustic crossing
models is disfavored by $\Delta\chi^2 \sim 4$ with respect to the
best parallax model. This is not surprising given the relatively weak
breaking of the continuous parallax symmetry shown in
Figures~\ref{fig-piE_cont} and \ref{fig-piEgrid}. Most of the parallax
signal comes from the acceleration of the Earth, and this can
be mimicked by xallarap with a large range of source star orbit
orientations. The small part of parameter space with $\Delta\chi^2 < 2$,
shown as black, maroon, or red in Figure~\ref{fig-xal_grid} corresponds
to $\lambda_s$ and $\beta_s$ very close to the ecliptic coordinates of 
the MOA-2007-BLG-192 source star. This indicates that the data do prefer
a source star orientation that allows it to mimic the orbital motion of the
Earth. This is evidence in favor of the microlensing parallax interpretation
because this is unlikely to occur by chance.

A small part of parameter space, with the Sun
in the plane of the source star orbit and orbital acceleration nearly
along the line-of-sight at peak magnification is disfavored much more
strongly than most of $\lambda_s$-$\beta_s$ space. These are geometries 
for which
acceleration during the event nearly vanishes, so that the observed
light curve effects cannot be reproduced.

\subsection{Constraints on Xallarap Orbits}
\label{sec-constrain_xal}

\begin{figure}
\plottwo{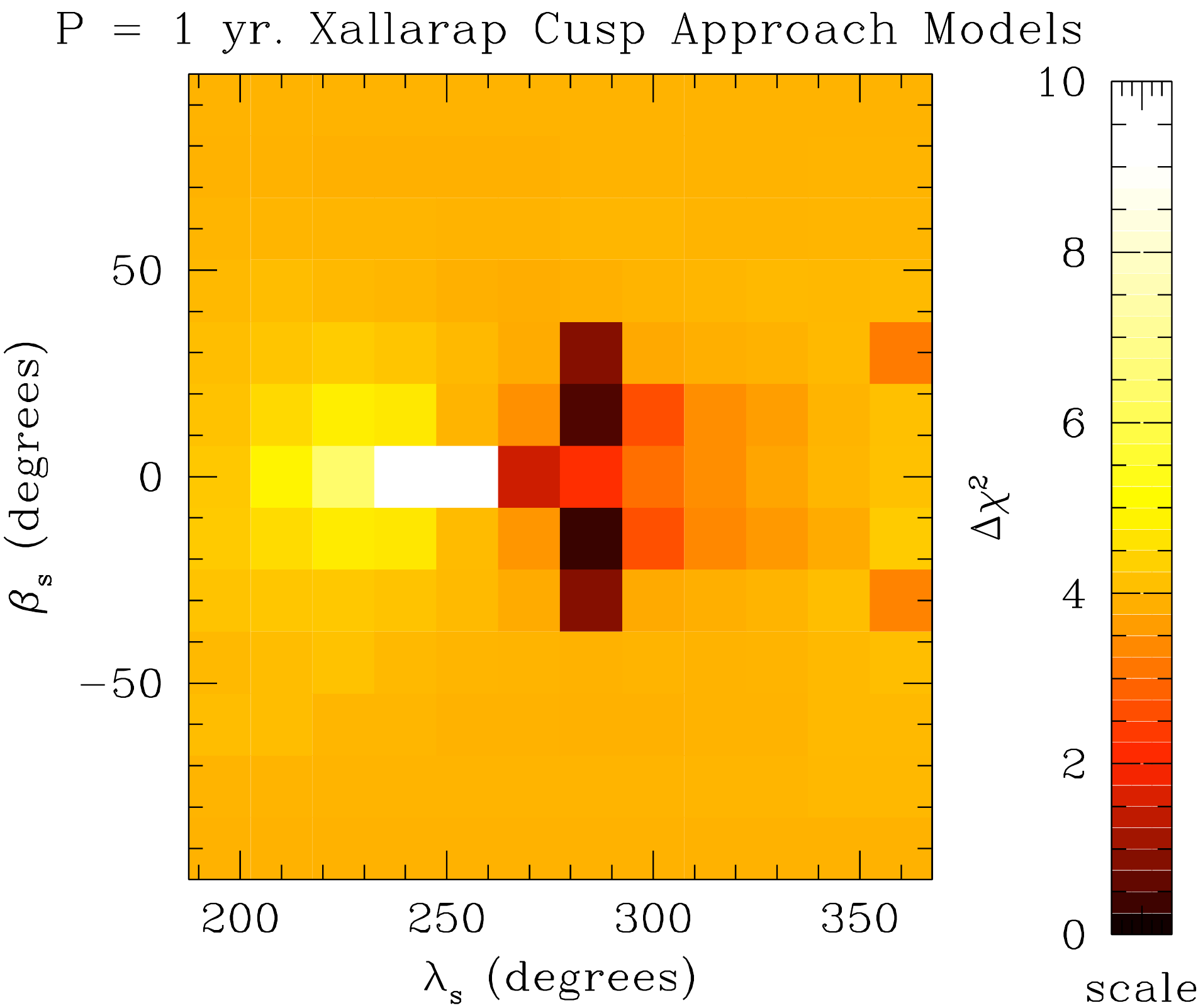}{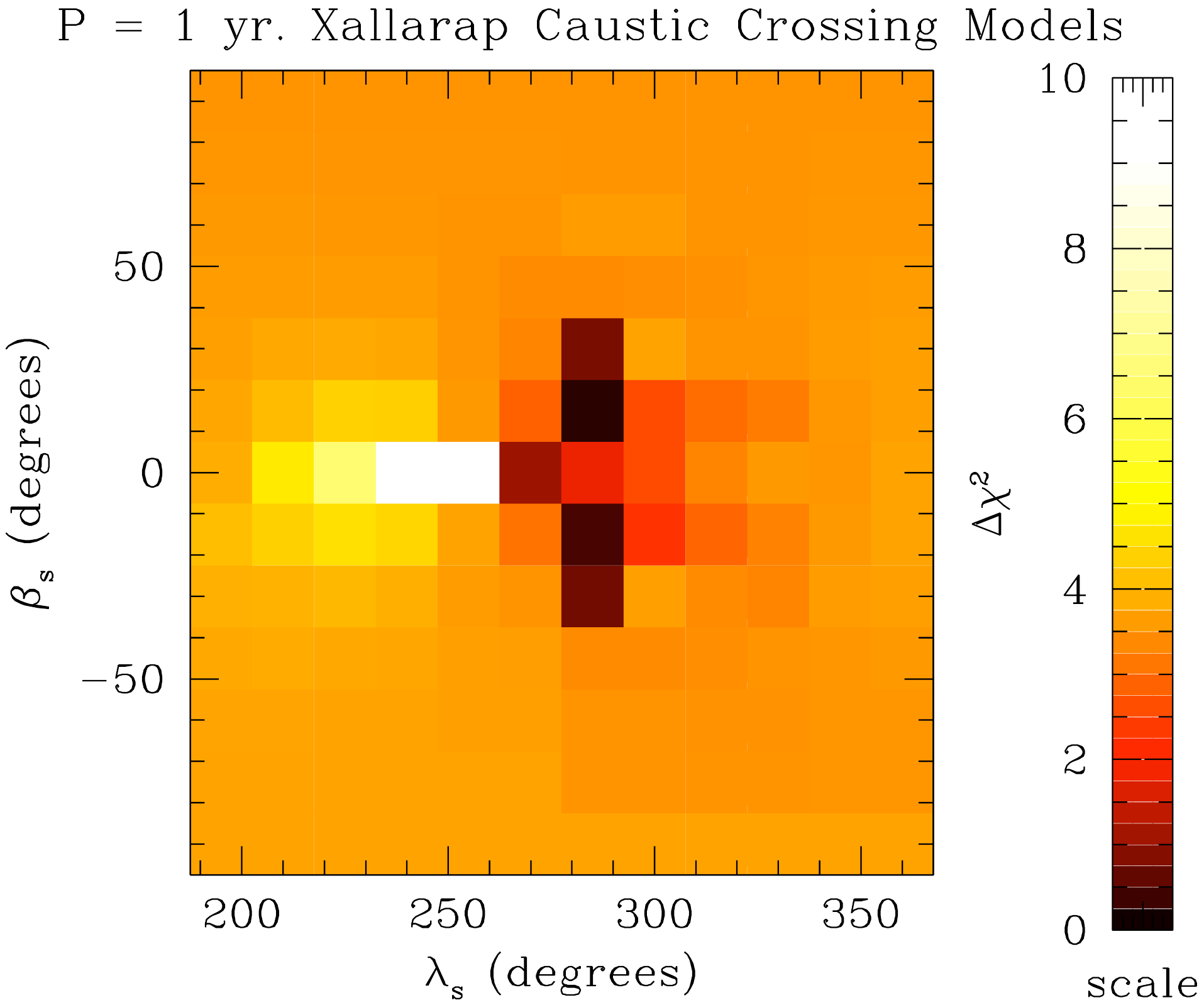}
\caption{The $\Delta\chi^2$ values between xallarap models
with the period fixed at $P=1\,$yr and the best microlensing parallax
models are plotted as a function of the pseudo-ecliptic longitude,
$\lambda_s$, and latitude, $\beta_s$ for both the cusp approach and
caustic crossing solutions.
\label{fig-xal_grid}}
\end{figure}

The models represented in Figure~\ref{fig-xal_grid} implicitly
assume that it is possible for the source star to have any
circular orbit. This would be true if it was reasonable to 
consider neutron stars or
black holes of all masses as possible binary companions
to the source star. However, neutron stars and black holes are 
quite rare compared to main sequence stars, and they are 
formed by processes that are likely to disrupt a binary system.
Certainly, some binary systems consisting of a black hole
and a low-mass main sequence star are observed as
x-ray binaries \citep{bh_araa} with periods of $\simlt 1\,$month.
The formation of such systems is not understood
\citep{bh_bin_form}, but they are extremely rare. So
if longer period black hole-main sequence
star binaries are as common as the x-ray binaries, the chances
are negligible that a black hole-main sequence star binary would
be the source for a microlensing event. A similar argument allows us
to reject the possibility of a neutron star companion to the source.

If we reject the possibility of a black hole or neutron star companion
to the source, we can consider only main sequence or white dwarf
source companions. We can constrain a possible main sequence star
companion with the upper limit on the combined brightness of lens and
source companion stars. A white dwarf is likely to be too faint to be detectable,
but we can constrain a possible white dwarf companion using the 
measured white dwarf mass function \citep{sloan_wd_mf}. Thus, we
have an upper limit on the mass of a companion to the source.

We can use Kepler's Third Law and the upper limit on the 
source companion mass, $M_C$, to constrain the magnitude of the xallarap 
vector. This constraint is
\begin{equation}
\xi_E \leq {M_C\over (M_C + M_S)^{2/3}\msun^{1/3}}{(P/{\rm yr})^{2/3}\over \theta_E D_S}
 \ ,
\label{eq-xal_lim}
\end{equation}
where $M_S$ is the source mass and $P$ is the orbital period. 
(This is the same as the constraint in
\citealt{dong_ogle71}, but the notation is somewhat different.) 
Note that this upper limit depends on the lower limit
on the angular Einstein radius, which in turn depends on the upper limit 
on the the source radius crossing time, $t_\ast$, through 
$\theta_E = \theta_\ast t_E/t_\ast$. So, this upper
limit is not affected by our weak lower limit on $t_\ast$.

\begin{figure}
\plottwo{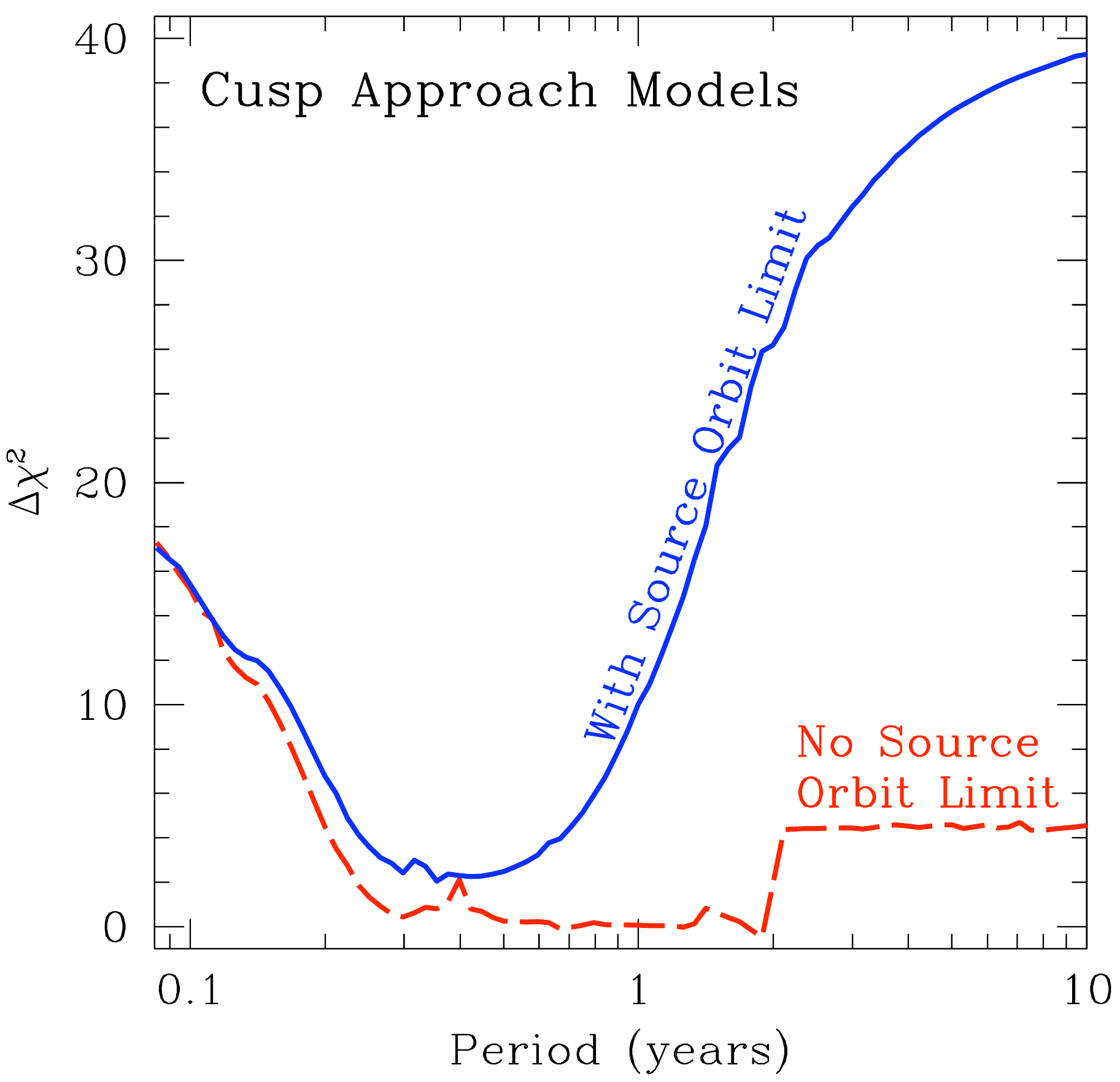}{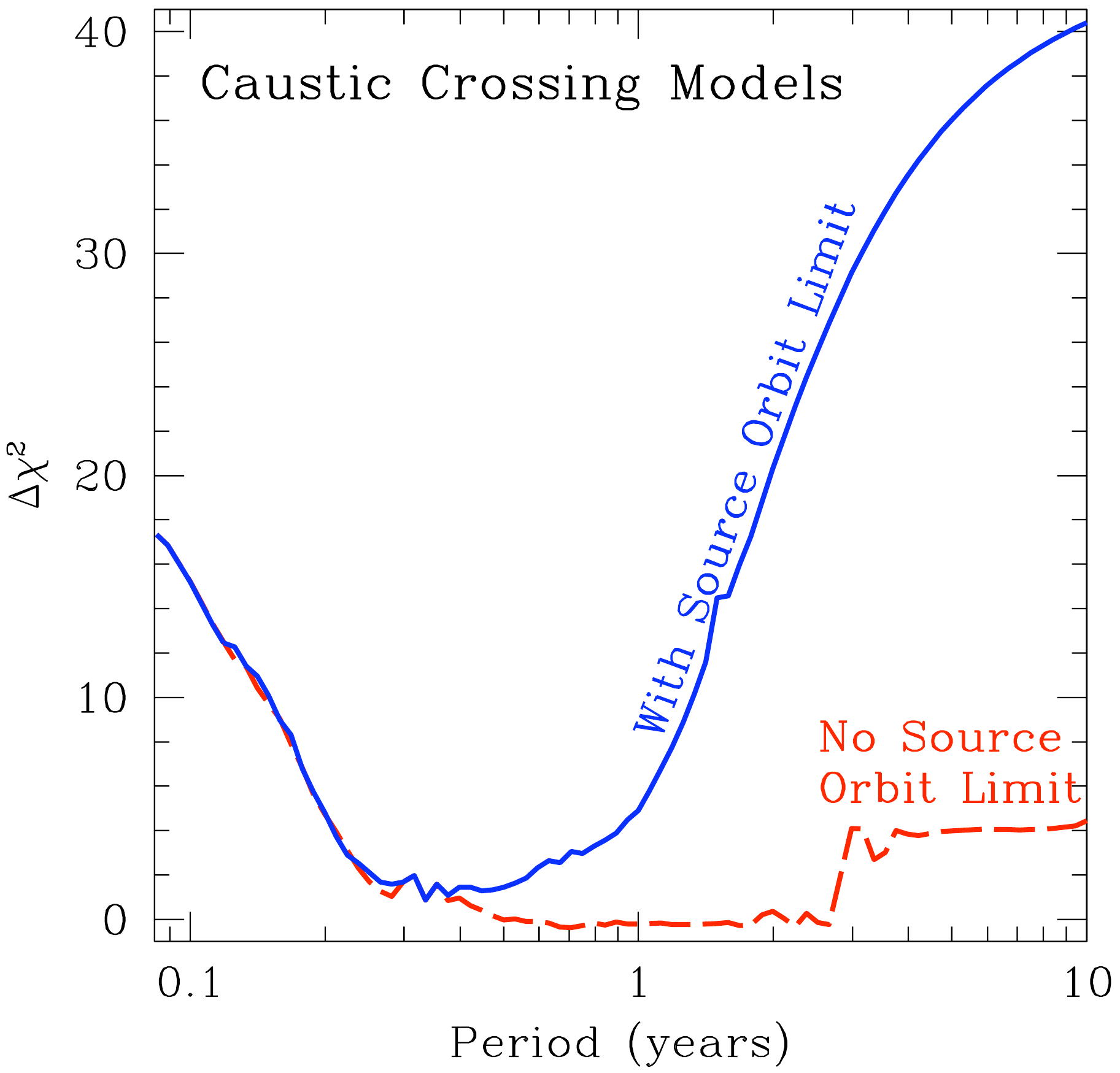}
\caption{The $\chi^2$ difference between the best xallarap and
microlensing parallax fits is plotted as a function of the 
orbital period of the source star and its companion for both
the cusp approach and caustic crossing models. The
red dashed curves in each figure indicate the best
fit with no constraint on the properties of the companion,
while the solid blue curves require that the source companion
have a mass of less than $0.7\,\msun$.
\label{fig-xalper}}
\end{figure}

A complete analysis of possible xallarap models would include both the 
possibility of a bright companion to the source as well as the flux from the
planetary host star superimposed on the flux of the source star, as well as
a full exploration of the possible xallarap parameter space. Such an analysis
would be quite complicated, and it would also not be definitive because we anticipate
additional VLT AO imaging and HST follow-up observations in the coming 
months and years. These follow-up data should provide tighter constraints
on the magnitude and colors of the source, lens, and possible source
companion. Therefore, we present a simplified analysis here. The solid
blue curves in Figure~\ref{fig-xalper} represent the $\chi^2$ values for
the best fit xallarap models as a function of source star period with 
an additional constraint contribution to $\chi^2$ given by
\begin{equation}
\chi^2_{\rm orb} = \Theta({\xi_{E,{\rm max}} -\xi_E})
\left({\xi_{E,{\rm max}} -\xi_E\over \sigma_\xi}\right)^2 \ ,
\label{eq-chi_xal}
\end{equation}
where $ \Theta$ is the step function, and the uncertainty, $\sigma_\xi$, in 
the maximum xallarap vector magnitude, $\xi_{E,{\rm max}}$, is taken to
be 20\% due to the uncertainty in $\theta_\ast$. (The uncertainties in
$t_\ast$ and $t_E$ are not included here because these are fit variables
that are adjusted to minimize the overall $\chi^2$.) We assume 
$D_S = 7.7\,$kpc, $M_S = 0.79\,\msun$, and $M_C = 0.7\,\msun$. This
$M_C$ value is taken to be an upper limit based upon the upper limit on
the brightness of the source and the white dwarf mass function.

Figure~\ref{fig-xalper} clearly indicates that xallarap models with
source star orbital periods of $\simgt 1\,$yr or $\simlt 0.25\,$yr are disfavored.
The constraint is somewhat weaker for the caustic crossing models because they
have larger values for $t_\ast$, so they imply a smaller $\theta_E$ than the 
cusp approach models do. The constrained xallarap models can be considered 
to have $\sim 1.5$ more degrees of freedom than the parallax model 
because they have two additional parameters ($\lambda_s$ and $\beta_s$) 
but also the constraint, eq.~(\ref{eq-chi_xal}) (which can be considered a ``half"
constraint because of the $\Theta$ function.)
Nevertheless, the $\chi^2$ values for the xallarap fits are worse than the
parallax fit $\chi^2$. The best cusp approach constrained xallarap fit has
$P = 0.422\,$yrs and $\Delta\chi^2 = 2.25$, and the best caustic crossing
constrained xallarap fit has $P = 0.376\,$yrs and $\Delta\chi^2 = 1.08$.
So, parallax is clearly favored, but not by a statistically significant margin,
based upon the fit $\chi^2$ values alone.

The {\it a priori} probability that the source star has a binary companion 
with a mass near $0.7\,\msun$ and a period of 0.25-1 year are relatively small.
The properties and prevalence of binary star systems have been studied by
\citet{bin_gdwarf} and \citet{lada_single}, and we can use these results for an
estimate. Since the source appears to be a K-dwarf, we can assume that
it has about a 50\% chance of having a binary companion. It is reasonable
to assume that the secondary stars were originally drawn from the same
initial mass function that applies to single stars and primaries 
\citep{bin_gdwarf}, and so we can estimate
the number of white dwarf secondaries by assuming that stars more
massive than the Sun have become white dwarfs. If we require
that the secondary must have a mass in the range
0.4-0.7$\,\msun$ to provide a sufficiently strong xallarap signal, then
we find that about 20\% of stars will have companion in the appropriate
mass range. However, most of these will not have orbital periods in the
range to provide the observed xallarap signal.

We can account for the binary period distribution and the somewhat disfavored
xallarap fits by summing over the results of the constrained fits shown in
Figure~\ref{fig-xalper}, with each fit weighted by $e^{-\Delta\chi^2/2}$ times
the {\it a priori} probability of a binary with the specified period, $P$.
The {\it a priori} probability distribution for $P$ is taken to be an equal
probability per $\log(P)$ for $1\,{\rm day} \leq P \leq 10^9\,$days. This is
slightly different from the distribution presented by \citet{bin_gdwarf}, but it
gives the correct probability in the period range of interest. This procedure
yields a xallarap probability of 0.30\% for the cusp approach models and
0.57\% for the caustic crossing models. With our assumed {\it a priori}
probability favoring the cusp approach models by 10:1 over the 
caustic crossing models, this gives a final xallarap probability of
0.32\%.

This indicates that xallarap is clearly disfavored. However, there are several
caveats to this analysis. The main one is that the stellar density in the bulge
is an order of magnitude larger than in the local disk. On the other hand, a
lens in the bulge would be much more massive than a disk lens, and it must
obey the upper limit on the excess flux at the position of the source. This
tends to rule a lens star in the densest part of the bulge.
If a source star companion saturates this limit, then the lens star cannot also
make a significant contribution. This limit on the brightness of the lens star
also tends to exclude the lens from the densest part of the bulge because the
implied lens mass grows to be quite large in the bulge. Thus, these additional
considerations might improve the probability of xallarap somewhat, but we
believe that xallarap is excluded at better than 98\% confidence.

\clearpage


\clearpage


\clearpage

\begin{deluxetable}{cccccccccccc}
\tablecaption{Planetary fit parameters \label{tab-fitpar} }
\tablewidth{0pt}
\tablehead{
\colhead{Fit}  & \colhead{$\chi^2$} & \colhead{$t_E$} & \colhead{$t_0^\prime$} &
       \colhead{$u_0$} &  \colhead{$d$} & \colhead{$\theta$} & \colhead{$q$} &
 \colhead{$t_\ast$} & \colhead{$I_s$} & \colhead{$\pi_E$} & \colhead{$\phi_E$}
}  

\startdata

A  & 1115.52 & 75.0 & 5.453 & -0.00364 & 0.881 & $113.6^\circ$ & $1.5\times 10^{-4}$ & 0.067 & 
        21.49 & 1.54 & $210.9^\circ$ \\
B  & 1115.46 & 74.5 & 5.453 & -0.00360 & 1.120 & $115.8^\circ$ & $1.2\times 10^{-4}$ & 0.064 & 
        21.48 & 1.52 & $211.7^\circ$ \\
C  & 1116.32 & 73.4 & 5.449 & 0.00381 & 0.874 & $246.8^\circ$ & $1.7\times 10^{-4}$ & 0.062 & 
        21.47 & 1.43 & $211.7^\circ$ \\
D  & 1116.37 & 73.8 & 5.448 & 0.00372 & 1.120 & $244.4^\circ$ & $1.2\times 10^{-4}$ & 0.063 & 
        21.48 & 1.53 & $211.0^\circ$ \\
E  & 1117.69 & 70.6 & 5.454 & -0.00390 & 0.879 & $113.9^\circ$ & $1.6\times 10^{-4}$ & 0.063 & 
        21.43 & 1.17 & $332.4^\circ$ \\
F  & 1117.84 & 69.1 & 5.451 & -0.00420 & 1.152 & $110.1^\circ$ & $2.3\times 10^{-4}$ & 0.068 & 
        21.40 & 1.20 & $332.3^\circ$ \\
G  & 1117.38 & 70.1 & 5.454 & 0.00395 & 0.885 & $246.8^\circ$ & $1.7\times 10^{-4}$ & 0.069 & 
        21.41 & 1.29 & $334.0^\circ$ \\
H  & 1117.35 & 69.9 & 5.456 & 0.00404 & 1.134 & $247.7^\circ$ & $1.9\times 10^{-4}$ & 0.069 & 
        21.41 & 1.21 & $331.6^\circ$ \\[0.25cm]
I  & 1115.14 & 75.1 & 5.462 & -0.00433 & 0.985 & $101.1^\circ$ & $2.1\times 10^{-4}$ & 0.117 & 
        21.49 & 1.60 & $211.7^\circ$ \\
J  & 1115.12 & 74.9 & 5.458 & -0.00420 & 1.007 & $103.8^\circ$ & $1.6\times 10^{-4}$ & 0.114 & 
        21.49 & 1.59 & $211.3^\circ$ \\
K  & 1115.59 & 69.4 & 5.455 & 0.00490 & 0.984 & $261.1^\circ$ & $2.4\times 10^{-4}$ & 0.117 & 
        21.41 & 1.50 & $213.6^\circ$ \\
L  & 1115.88 & 72.4 & 5.453 & 0.00442 & 1.006 & $256.1^\circ$ & $1.6\times 10^{-4}$ & 0.111 & 
        21.46 & 1.52 & $211.9^\circ$ \\
M  & 1116.81 & 68.7 & 5.459 & -0.00483 & 0.985 & $100.4^\circ$ & $2.3\times 10^{-4}$ & 0.116 & 
        21.39 & 1.30 & $332.7^\circ$ \\
N  & 1117.59 & 69.3 & 5.454 & -0.00452 & 1.005 & $105.0^\circ$ & $1.5\times 10^{-4}$ & 0.111 & 
        21.41 & 1.23 & $332.3^\circ$ \\
O  & 1116.36 & 68.9 & 5.463 & 0.00476 & 0.985 & $259.1^\circ$ & $2.3\times 10^{-4}$ & 0.117 & 
        21.39 & 1.34 & $333.6^\circ$ \\
P  & 1117.26 & 66.4 & 5.458 & 0.00475 & 1.006 & $255.8^\circ$ & $1.7\times 10^{-4}$ & 0.113 & 
        21.35 & 1.29 & $331.6^\circ$ \\
\enddata
\tablecomments{ This table shows the fit parameters for the 16 distinct planetary
models for MOA-2007-BLG-192. $t_0^\prime = t_0 -4240\,$days. $t_0$ and $u_0$
are the time and distance of the closest approach of the source to the lens center-of-mass.
$q$ and $d$ are the planet:star mass ratio and separation, and $\theta$ is the angle
between the source trajectory and the planet-star axis. $I_s$ is the best fit source
magnitude, and $\pi_E$ and $\phi_E$ are the magnitude and angle of the microlensing
parallax vector. The units for the Einstein radius crossing time,
$t_E$, the source radius crossing time, $t_\ast$, and $t_0^\prime$ are days, and all
other parameters
are dimensionless.}
\end{deluxetable}

\begin{deluxetable}{ccc}
\tablecaption{Parameter Values and MCMC Uncertainties - without prior
                          \label{tab-puncert} }
\tablewidth{0pt}
\tablehead{
\colhead{parameter}  & \colhead{value} & \colhead{2-$\sigma$ range} 
}  

\startdata

$M$  & $0.039{+0.022\atop -0.012} \msun$ & 0.019--$0.113\msun$ \\[0.15cm]
$m$  & $2.3{+2.5\atop -0.8} \mearth$ & 0.9--$14.4\mearth$ \\[0.15cm]
$a_\perp$  & $0.58{+0.18\atop -0.14} \,$AU & 0.32--$1.04\,$AU \\[0.15cm]
$D_L$ & $1.3\pm 0.5\,$kpc & 0.5--$2.3\,$kpc \\[0.15cm]
$I_S$ &  $21.44\pm 0.08$ & 21.30--21.60 \\[0.15cm]
$q$ & $1.9 \pm 0.8 \times 10^{-4}$ & 0.6--$6.4\times 10^{-4}$ \\[0.15cm]
\enddata
\end{deluxetable}

\begin{deluxetable}{ccc}
\tablecaption{Parameter Values and MCMC Uncertainties - with prior
                         \label{tab-ppuncert} }
\tablewidth{0pt}
\tablehead{
\colhead{parameter}  & \colhead{value} & \colhead{2-$\sigma$ range} 
}  

\startdata

$M$  & $0.060{+0.028\atop -0.021} \msun$ & 0.024--$0.128\msun$ \\[0.15cm]
$m$  & $3.3{+4.9\atop -1.6} \mearth$ & 1.0--$17.8\mearth$ \\[0.15cm]
$a_\perp$  & $0.62{+0.22\atop -0.16} \,$AU & 0.33--$1.14\,$AU \\[0.15cm]
$D_L$ & $1.0\pm 0.4\,$kpc & 0.5--$2.0\,$kpc \\[0.15cm]
$I_S$ &  $21.44\pm 0.08$ & 21.31--21.61 \\[0.15cm]
$q$ & $1.8 {+1.9\atop -0.8} \times 10^{-4}$ & 0.5--$7.1\times 10^{-4}$ \\
\enddata
%
%
\end{deluxetable}

\end{document}